\documentclass{article}
\usepackage[round]{natbib}
\usepackage[english]{babel}
\usepackage{amsmath}
\usepackage{physics}
\usepackage{amssymb}
\usepackage{amsthm}
\usepackage{commath} 
\usepackage{eufrak}
\usepackage{mathtools}
\usepackage{physics}
\usepackage{epsfig}
\usepackage{graphics}
\usepackage{bm}
\usepackage{tikz}
\usetikzlibrary{arrows,positioning,shapes.geometric,decorations.markings}
\tikzstyle{squarenode}=[rectangle,draw]
\tikzstyle{littlenode}=[circle,draw,minimum size=0.5cm,font=\small]
\tikzstyle{bigellipse}=[ellipse,draw,x radius=10cm, y radius=5cm]
\tikzset{negate/.style={
            decoration={markings,
            mark= at position 0.5 with {
                  \node[transform shape] (tempnode) {$\Big\Vert$};
                  }
              },
              postaction={decorate}
}
}
\usepackage{xcolor}
\usepackage{tabularx}
\usepackage{multirow}
\usepackage{url}
\usepackage{booktabs}
\usepackage{rotating}
\usepackage{caption}
\usepackage{floatrow}
\usepackage{graphicx} 
\usepackage{graphics} 
\usepackage{caption}
\usepackage{subfig}
\usepackage{pgfplots}
\usepackage{authblk}
\usepackage{etoolbox}

\pgfplotsset{compat=1.11}
\RequirePackage{hyperref}

\usepackage{marvosym}
\usepackage[marvosym]{tikzsymbols}

\usetikzlibrary{arrows,backgrounds,automata,positioning}
\usetikzlibrary{shapes.geometric,decorations.markings}
\usetikzlibrary{decorations.text}
\usetikzlibrary{decorations.pathmorphing}
\tikzstyle{mynode}=[circle,draw,minimum size=0.8cm]
\tikzstyle{myemptynode}=[circle,minimum size=0.8cm]
\tikzstyle{squarenode}=[rectangle,draw]
\tikzstyle{little}=[circle,draw,fill=black,minimum size=0.15cm,inner sep=0pt]
\tikzstyle{littlehollow}=[circle,draw,minimum size=0.15cm,inner sep=0pt]
\tikzset{negate/.style={
            decoration={markings,
            mark= at position 0.5 with {
                  \node[transform shape] (tempnode) {$||$};
                  }
              },
              postaction={decorate}
}
}

\hypersetup{
	colorlinks=true,
	linkcolor=black,
	citecolor=black,
	filecolor=black,
	urlcolor=black,
}

\newcolumntype{Y}{>{\centering\arraybackslash}X}
\newcolumntype{M}{>{\centering\arraybackslash}m}

\renewcommand{\rm}[1]{\mathrm{#1}}
\theoremstyle{plain}

\makeatletter
\renewcommand*\env@matrix[1][\arraystretch]{%
  \edef\arraystretch{#1}%
  \hskip -\arraycolsep
  \let\@ifnextchar\new@ifnextchar
  \array{*\c@MaxMatrixCols c}}
\makeatother

\usepackage{soul}
\newcommand{\tu}{\textup}

\title{Modeling within-department homogeneity in research quality rankings: an application to the Italian ISPD}

\author[1]{Giorgio E. Montanari}
\author[2]{Marco Doretti\thanks{email: marco.doretti@unifi.it}}

\affil[1]{Department of Political Science, University of Perugia}
\affil[2]{Department of Statistics, Computer Science, and Applications, University of Florence}

\date{}

\begin{document}
\maketitle

\begin{abstract}
In this paper, we consider the academic department ranking system of Italy, which is based on a performance index named \emph{Indice Standardizzato di Performance Dipartimentale} (ISPD). While critiques to the ISPD have been moved for its marked tendency to polarization, we here formalize a yet unexplored determinant of this phenomenon, that is, the presence of within-department homogeneity among the standardized scores used to build the index. We account for this intra-departmental correlation by modeling it as a function of departments' size. The proposed model, estimated via Maximum Likelihood, allows to build a fairer ranking procedure via the definition of a properly adjusted version of the ISPD. The estimation framework is also adapted to fit publicly available data, which are coarsened by rounding and/or left-truncated. To this end, a novel probability distribution termed \emph{Betoidal} is introduced. Empirical evidence in favor of the proposed model is found in the 2017 and 2022 data. Moreover, a simulation study shows that the adjusted index significantly overcomes not only the original ISPD, but also other more data-demanding competing proposals.
\end{abstract}


\section{Introduction}\label{sec:intro}

Many countries around the world have developed evaluation systems of their universities and high research institutions in order to stimulate improvements in the quality of teaching and scientific research, and the dissemination and utilization of research results. These systems are largely based on the assessment of institutional activities, and their results are used to guide the allocation of public resources and to introduce incentives and penalties, thus encouraging universities and research institutions to improve their performance. Funding schemes based on evaluation tend to stimulate both the volume and quality of research, as well as to enhance interaction with industry, promote internationalization, and strengthen the socio-economic impact of academic research~\citep{GeunaMartin2003,Hicks2012}. However, the potential for unintended consequences arising from opportunistic behavior among evaluated institutions -  focused on improving measurement indicators rather than the quality of processes and services -must be carefully considered. Competition for resources may also occur at the expense of collaboration among institutions~\citep{JonkersZacharewicz2017}. Furthermore, the costs associated with establishing and maintaining assessment systems represent another critical issue~\citep{GeunaPiolatto2016}.

Evaluation and performance-based research funding systems vary considerably across countries. In Europe, several countries adopt funding schemes based on the evaluation of research outputs using quantitative bibliometric indicators, peer-review assessments, or both (for more details on national evaluation strategies see~\citealt{JonkersZacharewicz2017}). However, the specific design of funding formulae and evaluation criteria differs substantially, making it difficult to identify a single best practice~\citep{Zacharewiczetal2019}. This paper focuses on the case of Italy, where universities are primarily financed through state contributions provided by the Ordinary Financing Fund (in Italian, {\em Fondo per il Finanziamento Ordinario delle Università} - FFO). Nowadays, approximately 30\% of the total FFO allocation is distributed according to performance-based criteria, and 60\% of this amount depends on research quality. A dedicated agency (\emph{Agenzia Nazionale per la Valutazione del Sistema Universitario e della Ricerca}  - ANVUR) is entrusted by the Ministry of University and Research with conducting such a quality assessment exercise, named {\em Valutazione della Qualità della Ricerca} (VQR).

Results from the VQR exercise are also used to award additional funds (on the basis of the ranking position and a specific department development project) to a subset of the top-performing 350 academic departments, labeled {\em Departments of Excellence}. To identify top-350 list, ANVUR produces, on a five-year basis, a ranking based on a single quantitative measure, the Standardized Index of Department Performance (in Italian, \emph{Indice Standardizzato di Performance Dipartimentale} - ISPD); see~\cite{PoggiNappi2014} and~\cite{Poggi2015}. Such an index is a function of the standardized scores attributed by expert committees to a set of research outputs each department autonomously selects. The first ISPD-based ranking was developed in 2017~\citep{MUR2017} using results of the VQR exercise based on the university outputs in the years 2010-2014 (hereafter, VQR 2010-2014). It was used to identify the Departments of Excellence for the years 2018-2022. The second ranking was released in 2022~\citep{MUR2022} to identify the Departments of Excellence for the years 2023-2027; it was based on VQR 2015-2019. Next ranking is expected in 2027 to cover the period 2028-2032; it will rely on the ongoing VQR 2020-2024.

Several critiques have been moved to various aspects of the VQR evaluation framework; see~\cite{GalliGreco2025} and references therein for a recent review. In particular,~\cite{GalliGreco2025} indicate two major limitations of the ISPD: the fact that standardization of scores occurs within each Scientific Disciplinary Sector, and the overall index tendency toward polarization - that is, to assign extreme (minimum or maximum) values to a large number of departments. The latter reduces the discriminatory power of ISPD, limiting its ability to distinguish truly outstanding (deficient) departments from those that merely perform above (below) average. According to the authors, it is due to the use, within the ISPD formula, of the so-called scaled average, which corresponds to the average score obtained by each department multiplied by the square root of the number of the submitted research outputs. To mitigate these issues, an alternative version of the ISPD has been proposed. 

In this study, we review the whole process leading to the construction of the ISPD and identify a specific feature which is, to the best of our judgment, indeed responsible for its observed polarization: the higher homogeneity of research output scores within the same department (compared to the whole population of outputs). This homogeneity, not accounted for either in ANVUR's framework or in subsequent analyses, is mainly due to the following facts: \emph{i)} each scholar typically delivers more than one research output, \emph{ii)} outputs are often produced by scholars affiliated to the same department, and \emph{iii)} departments are generally quite homogeneous with respect to the scientific capabilities of their components. From a statistical standpoint, homogeneity induces correlation among the research output scores of the same department. We handle the presence of this intra-departmental correlation by modeling it as a function of departments' size. We suggest a consequently adjusted version of the ISPD that not only possesses significantly higher discriminatory power at the tails, but also enables fairer pairwise comparisons in the middle of the distribution. Further, we show how the proposed correlation model can be estimated from publicly available data for the ANVUR 2017 and 2022 ranking exercises. To this end, we present a novel probability distribution that resembles the symmetric Beta and is thus termed \emph{Betoidal}.

The remainder of the paper is organized as follows. Section~\ref{sec:background} describes the original ANVUR approach (Section~\ref{subsec:anvur}) as well as the effect of intra-departmental correlation on both polarization and fairness of the ISPD (Section~\ref{subsec:rho}). Section~\ref{sec:correction} presents the proposed intra-departmental correlation model and its Maximum Likelihood (ML) estimation. It also shows how the adjusted ISPD can be obtained. Section~\ref{sec:estpubl} addresses estimation with publicly available data. In detail, Section~\ref{subsec:betoidal} introduces the Betoidal distribution and its left-truncated version, which are essential to estimate the intra-departmental correlation model based on ANUVR 2017 and 2022 data. The ML specifics of these two cases are reported in Sections~\ref{subsec:coarse} and~\ref{subsec:coarsetrunc}, respectively, whereas estimation results are summarized in Section~\ref{subsec:res}. These results permit to calibrate the simulation study presented in Section~\ref{sec:simulazio}, where the behavior of the adjusted ISPD is compared to that of the original ANVUR ISPD and of other competing proposals in a setting very close to the one of the ANVUR 2017 exercise. Finally, Section~\ref{sec:concl} summarizes the main findings and offers some concluding remarks.


\section{Background}\label{sec:background}

\subsection{The ANVUR approach for building the ISPD index}\label{subsec:anvur}

At the beginning of every VQR exercise, each department $d=1,\dots,D$ is required  to submit to ANVUR a selection of $N_d$ Scientific Products (SPs), where $N_d$ equals $k$ times the number of affiliated scholars (net of minor exceptions). In detail, $k$ was set to 2 in the VQR 2010-2014,  to 3 in the VQR 2015-2019, and to 2.5 in the ongoing VQR 2020-2024. SPs are assigned a score by expert committees through informed peer-review. The admissible scores are 0, 0.2, 0.5, 0.8 or 1, with higher values corresponding to more favorable evaluations. Since SPs are classified into Scientific Disciplinary Sectors (SDSs), each score is standardized with the mean and the standard deviation of its SDS before being further processed. This procedure aims to mitigate potential discrepancies arising from heterogeneous evaluation practices across scientific fields as well as from differences in the standards adopted by different committees. A critical discussion of this standardization can be found in~\cite{GalliGreco2025}. In any case, it is worth to emphasize that such a process effectively removes between-SDS differences in average research quality: department excellence is therefore interpreted as the ability to present SPs that perform above the average of their respective SDSs.

Denoting by $z_{di}$ the standardized score of the $i$-th SP ($i=1,\dots,N_{d}$) submitted by department $d$, the ANVUR ranking procedure is based on the transformation $x_d=\Phi(\tilde{z}_d)$, where $\Phi(\cdot)$ is the standard Normal Cumulative Distribution Function (CDF) and $\tilde{z}_d=\sqrt{N_d}\bar{z}_d$ is the so-called scaled average of the department, with 
\[
\bar{z}_d\equiv \frac{\sum_{i=1}^{N_{d}} z_{di}}{N_d}
\]
denoting the mean of the standardized scores of the submitted SPs. The transformed value $x_d$ is expressed as a percentage and rounded to the nearest  half-integer to obtain the final ISPD for department $d$. Formally,
\begin{equation}\label{GEM1}
\textup{ISPD}_d=\lfloor 200\cdot x_d + 0.5 \rfloor/2.
\end{equation}

The rationale of the ISPD consists in the fact that each department $d$ is associated to a Virtual Department (VD) such that the score of every SP submitted by $d$ is replaced by one drawn at random and independently from the population of SP scores belonging to the same SDS and submitted by all $D$ departments under evaluation. In this way, department $d$ can be fairly compared to its randomly generated equivalent, $\text{VD}_d$, and the transformed scaled average $x_d$ can be interpreted as the probability that the average score of $\text{VD}_d$ does not exceed $\bar{z}_d$. This construction provides an effective device to account for the heterogeneity of Italian departments in terms of their SDS composition. 

From a probabilistic perspective, the SP scores of $\text{VD}_d$ can be regarded as realizations of discrete standardized random variables $Z_{di}$. Since $N_d$ typically exceeds 60 (with a few exceptions), under the assumption of mutual independence among these variables, the Central Limit Theorem (CLT) justifies treating VD scaled averages as realizations of random variables, $\tilde{Z}_d$, approximately distributed as standard Normal ones, i.e., $\tilde{Z}_d\approx N(0,1)$. Consequently, the distribution of $\tilde{Z}_d$ is (approximately) the same for all departments, which ensures the comparability of the ISPDs across departments and the effectiveness of the resulting ranking. In other words, the use of scaled averages removes dependence of ISPDs on the sample size, thereby avoiding the well-known tendency of the ``traditional'' mean to place larger (smaller) institutions in central (extreme) ranking positions~(\citealt{GoldsteinSpiegelhalter1996}; see also \citealt{Normandetal2016} for a related discussion in the context of hospitals' evaluation).
  
As noted above, $x_d$ can be interpreted as the probability that a randomly generated VD with the same SDS structure as $d$ performs no better than $d$ in terms of the mean score $\bar{z}_d$. Ideally, this probability should behave as an order statistic and hence follow (approximately) a uniform distribution on the unit interval. Indeed, letting $F_{X_d}(x)$ denote the CDF of the transformed random variable $X_d=\Phi(\tilde{Z}_d)$, at $x\in[0,1]$ we have
\[
\begin{split}
F_{X_d}(x) &= P(X_d \leq x) = P(\Phi(\tilde{Z}_d) \leq x) = P(\tilde{Z}_d \leq \Phi^{-1}(x)) \\
&\approx \Phi(\Phi^{-1}(x)) = x,
\end{split}
\]
where $\Phi^{-1}(\cdot)$ denotes the standard Normal quantile function. 


\subsection{The impact of intra-department correlation}\label{subsec:rho}

\begin{figure}[tb]
\centering{
\subfloat[][\label{fig:ispd17}]{\includegraphics[scale=0.53]{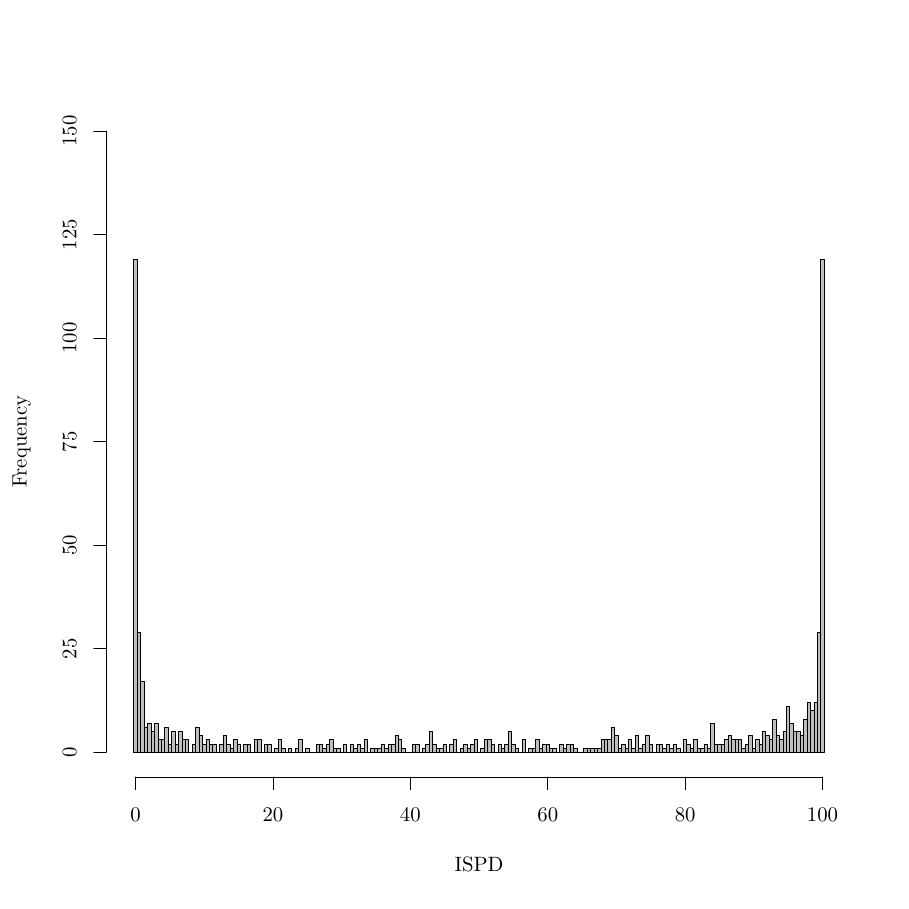}}\quad
\subfloat[][\label{fig:ispd22}]{\includegraphics[scale=0.53]{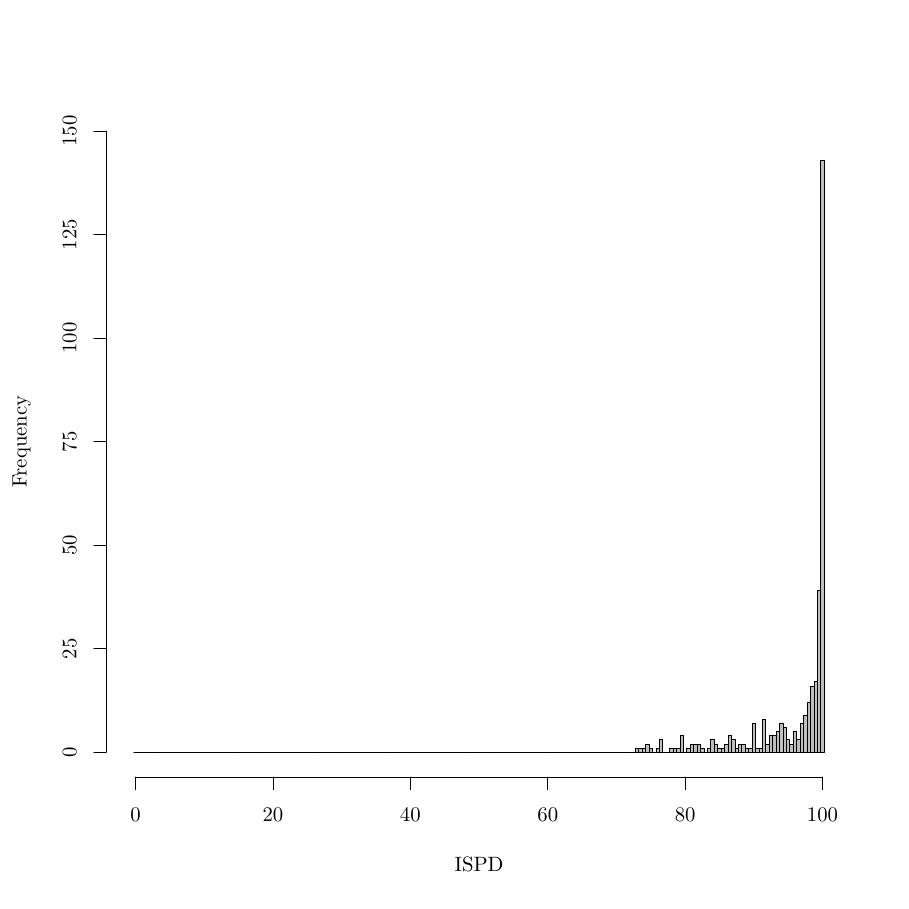}}
}
\caption{Distribution of ISPDs for ANVUR's (a) 2017 ranking exercise (766 departments) and (b) 2022 exercise (top 350 departments only).}
\label{fig:ispd}
\end{figure}

The framework outlined above relies on the assumption that the untransformed VD scaled averages can be treated, for all departments, as realizations from an (approximately) standard Normal distribution. Under this assumption, the distribution of ISPDs across departments should be approximately uniform over the grid $[0, 0.5, 1,\dots, 99, 99.5, 100]$. However, empirical evidence points in a different direction. An independent Italian association engaged in the analysis of research evaluation systems documented that the ANVUR 2017 ranking exhibits a markedly U-shaped distribution~\citep{Roars}, with symmetric peaks at the lower and upper bounds of the ISPD scale (0 and 100; see Figure~\ref{fig:ispd17}). In other words, ISPD values appear to be highly concentrated at the extremes rather than evenly spread. A similar pattern, even more pronounced, emerges in the 2022 ranking~\citep{Roars}, although only the right tail of the distribution (i.e., the top-350 list Departments of Excellence are selected from) is publicly available for that exercise (Figure~\ref{fig:ispd22}).

The above evidence suggests that scaled averages of the departments are indeed observations from centered (approximate) normal distributions with standard deviations $\sigma_d>1$, reflecting within-department correlation between scores. Hence, the observed scaled average $\tilde{z}_d$ should not be compared to a VD constructed under the assumption of mutual independence between scores; instead, it should be compared to a VD randomly generated under the same correlation structure that characterizes department $d$. In this framework, letting $\mathcal{N}_{d}=\{1,\dots,N_d\}$, and $\mathcal{N}_{di}=\{1,\dots,N_d\}\setminus i$)
the variance of $\tilde{Z}_d$ becomes
\begin{equation}\label{eq:vrho}
\begin{split}
\sigma^2_d &\equiv V(\tilde{Z}_d) = \frac{1}{N_d}\Biggl[\sum_{i\in \mathcal{N}_d}V(Z_{di}) +\sum_{i\in\mathcal{N}_{d}}\sum_{i'\in\mathcal{N}_{di}}\textup{Cov}(Z_{di},Z_{di'}) \Biggr] \\
&=1+\rho_d(N_d-1),
\end{split}
\end{equation}
where 
\[
\rho_d=\frac{\sum_{i\in\mathcal{N}_{d}}\sum_{i'\in\mathcal{N}_{di}}\rho_{d,ii'}}{N_d(N_d-1)}
\]
denotes the average within-department correlation between pairs of SPs.

Expression~\eqref{eq:vrho} shows that, whenever $\rho_d>0$, the variance of the scaled average increases with department size $N_d$. This would hold even if average correlations were constant across departments (i.e., $\rho_d\equiv\rho$). As a consequence, scaled averages can no longer be interpreted as draws from a common distribution: their dispersion becomes size-dependent, thereby reintroducing the dimensionality effect that the scaling procedure was meant to eliminate. In particular, since $\sigma^2_d$ grows with $N_d$, larger institutions are more likely to be assigned extreme ISPD values; in this respect~\cite{GalliGreco2025} first noted this effects of department size. This dynamic mirrors, though in the opposite direction, the well-known critique of the ``traditional'' average discussed in Section~\ref{subsec:anvur}~\citep{GoldsteinSpiegelhalter1996,Normandetal2016}. Such an effect clearly reflects on the ISPD, which is a monotonic transformation of $\tilde{z}_d$.

\begin{figure}[tb]
\centering{
\subfloat[][\label{fig:ex3}]{\includegraphics[scale=0.53]{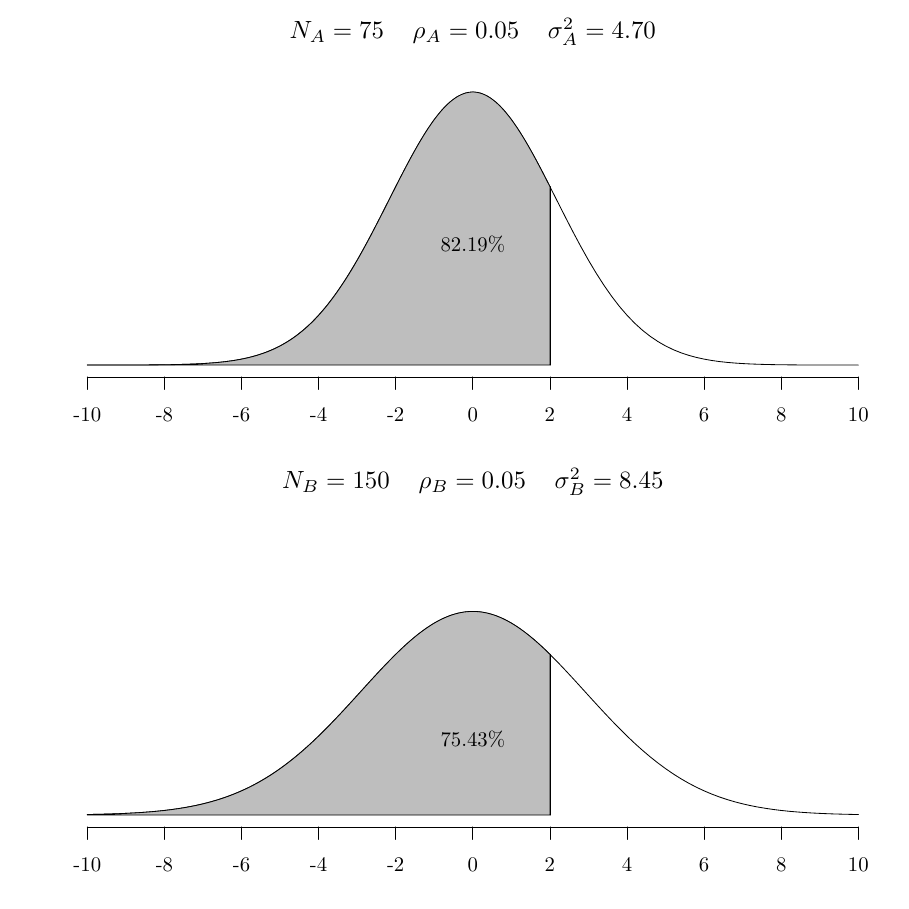}}\quad
\subfloat[][\label{fig:ex4}]{\includegraphics[scale=0.53]{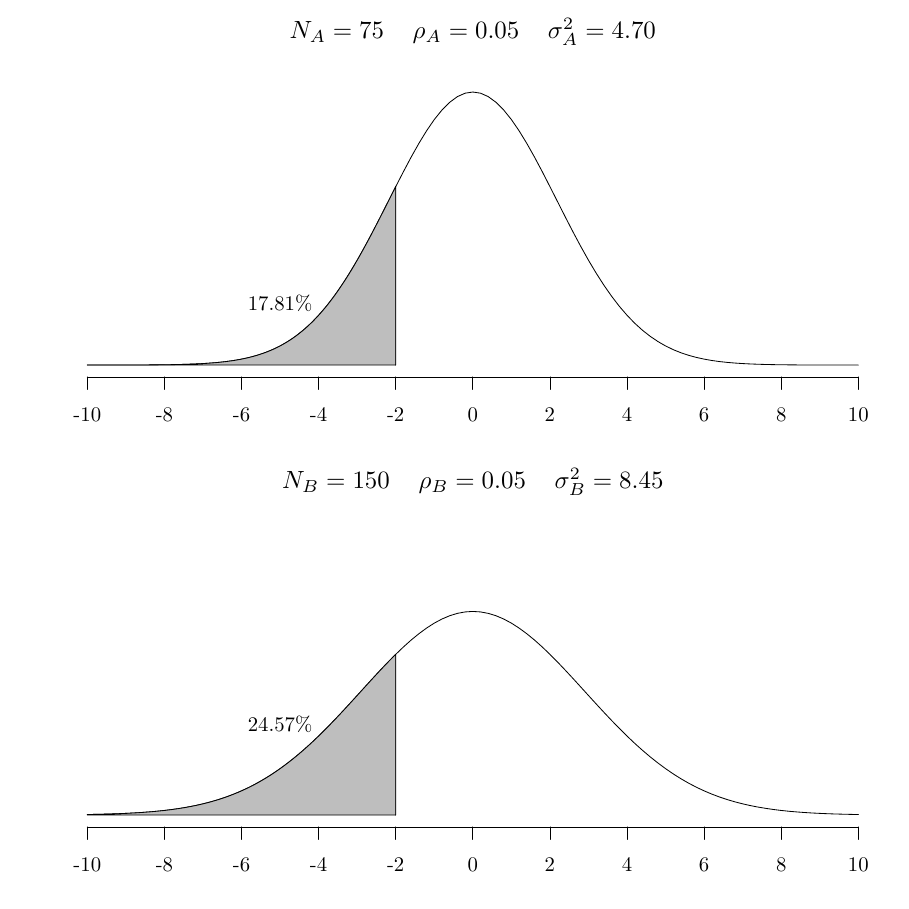}}
}
\caption{Two settings with $\rho_{\textup{A}}=0.05=\rho_{\textup{B}}$, $N_{\textup{A}}=75$, $N_{\textup{B}}=150$, and (a) $z_{\textup{A}}=2=z_{\textup{B}}$, and (b) $z_{\textup{A}}=-2=z_{\textup{B}}$. In (a), $\textup{ISPD}_{\textup{A}}=97.72\%=\textup{ISPD}_{\textup{B}}$, with the correct performance measures being $P(Z_{\textup{A}} \leq 2)=82.19\%$ and $P(Z_{\textup{B}} \leq 2)=75.43\%$. In (b), $\textup{ISPD}_{\textup{A}}=2.28\%=\textup{ISPD}_{\textup{B}}$, with $P(Z_{\textup{A}} \leq -2)=17.81\%$ and $P(Z_{\textup{B}} \leq -2)=24.57\%$.}
\label{fig:ex}
\end{figure}

Beyond producing polarization at the extremes of the ISPD range, the above mechanism also affects the fairness of comparisons. To see this, consider two departments, $\textup{A}$ and $\textup{B}$, that attain the same scaled average (and hence the same ISPD) and share the same intra-department correlation, but differ in size, with $N_{\textup{B}}>N_{\textup{A}}$. Since $\sigma^2_{\textup{B}}>\sigma^2_{\textup{A}}$, the observed scaled average is therefore statistically less extreme for Department $\textup{B}$ than for Department $\textup{A}$. This is illustrated in Figure~\ref{fig:ex}, where $\rho_{\textup{A}}=\rho_{\textup{B}}=0.05$, $N_{\textup{A}}=75$, $N_{\textup{B}}=150$, and (a) $z_{\textup{A}}=z_{\textup{B}}=2$, or (b) $z_{\textup{A}}=z_{\textup{B}}=-2$: both departments obtain the same ISPD and are pushed towards the boundaries of the 0--100 interval, but distortion is more pronounced for the larger department. Analogous distortions arise when departments share the same scaled average and size but differ in their intra-departmental correlation levels. In all these cases, the probabilistic interpretation underlying the ISPD no longer holds, and cross-department comparisons become unfair.

In summary, the current ISPD formulation does not fully achieve its intended goal of ensuring comparability across heterogeneous departments. Indeed, when intra-department correlation is present, the scaled average $\tilde{z}_d$ should be compared to that of a VD characterized by the same correlation level $\rho_d$. Only under this alignment does $\tilde{z}_d$ belong to the appropriate reference probability distribution. Also, to restore comparability across departments of different sizes, scaled averages should be standardized so that they can be interpreted as realizations from a common distribution. Without this correction, the ranking mechanism implicitly favors or penalizes departments depending on their size and internal dependence structure, thereby undermining fairness and distorting incentives in research assessment.


\section{The intra-departmental correlation model and ISPD adjustment}\label{sec:correction}

In light of the concepts discussed in Section~\ref{subsec:rho}, the most natural adjustment of the ISPD involves rescaling $\tilde{z}_d$ by its department-specific standard deviation. In detail, we have
\begin{equation}\label{eq:ispdtheo}
\textup{ISPD}_d^{\textsc{theo}} = \lfloor 200\cdot x_d^{\textsc{theo}} + 0.5 \rfloor/2, 
\end{equation}
where $x_d^{\textsc{theo}} = \Phi(\tilde{z}_d/\sigma_d)$, $\sigma_d = \{1+\rho_d(N_d-1)\}^{1/2}$, and the expression of $\sigma_d$ follows from~\eqref{eq:vrho}. 
This adjustment restores the interpretation of the final performance index as the percentile position of department $d$, relative to an appropriate reference distribution. From a policy perspective, it preserves the original rationale of ANVUR’s approach - namely, benchmarking departments against virtual counterparts - while correcting for a structural feature (intra-department dependence) that would otherwise generate artificial polarization and size-related distortions. The resulting index remains transparent, monotonic in performance, and easy to communicate. However, it is important to remark that it still invokes (approximate) Normality of the VD scaled average $\tilde{Z}_d$, although it is a function of correlated random variables. Under standard regularity conditions, this property holds by virtue of a suitably modified version of the CLT~\citep{DavydovIbragimov1971,HallHeyde1980,Chudiketal2011}, provided that the product $\rho_d N_d$ has a finite limit as $N_d$ tends to infinity (see Appendix~\ref{app:clt} for details).

Clearly, index in~\eqref{eq:ispdtheo} cannot be computed from observed data, since it involves the theoretical standard deviation $\sigma_d$, which is an unknown parameter. To obtain a feasible measure, such a quantity needs to be estimated, which in turn amounts to estimating $\rho_d$. When SP-level scores (also referred to as micro-data) are available, an unbiased non-parametric estimator of $\rho_d$ is given by the average cross-product of standardized scores. This leads to the estimate
\begin{equation}\label{eq:zrhomicro}
\hat{\rho}_d^{\textsc{np}} = \frac{1}{N_d(N_d-1)}\mathop{\sum_{i\in\mathcal{N}_{d}}}\sum_{i'\in\mathcal{N}_{di}}z_{di}z_{di'}
\end{equation}
and to the consequent correction
\begin{equation}\label{eq:ispdmicro}
\textup{ISPD}_d^{\textsc{np}} = \lfloor 200\cdot x_d^{\textsc{np}} + 0.5 \rfloor/2, 
\end{equation}
where $x_d^{\textsc{np}} = \Phi(\tilde{z}_d/\hat{\sigma}^{\textsc{np}}_d)$, and $\hat{\sigma}^{\textsc{np}}_d = \{1+\hat{\rho}^{\textsc{np}}_d(N_d-1)\}^{1/2}$.

However, as a second-order statistic, $\hat{\rho}_d^{\textsc{np}}$ may be unstable and occasionally fall outside the admissible range $[-(N_d-1)^{-1};1]$, especially for moderate $N_d$. A possibly more robust solution involves fitting a Random Intercept Model~\citep[RIM, ][]{Goldstein1986,Goldstein2011} on the standardized scores, with department membership taken as grouping factor. The intra-class correlation coefficient estimated from this model, $\hat{\rho}^{\textsc{rim}}$, could be then used to obtain the correction
\begin{equation}\label{eq:ispdrim}
\textup{ISPD}_d^{\textsc{rim}} = \lfloor 200\cdot x_d^{\textsc{rim}} + 0.5 \rfloor/2 
\end{equation}
where $x_d^{\textsc{rim}} = \Phi(\tilde{z}_d/\hat{\sigma}^{\textsc{rim}}_d)$, and $\hat{\sigma}^{\textsc{rim}}_d = \{1+\hat{\rho}^{\textsc{rim}}(N_d-1)\}^{1/2}$. The problem with this approach is that it implicitly assumes that intra-departmental correlation does not vary across departments; which is in contrast with the above mentioned condition for the applicability of the modified CLT.

An alternative strategy circumventing the above issues - and relying only on departmental scaled averages (i.e., macro-data) - consists in specifying a parsimonious model which links $\rho_d$ directly to the department's size $N_d$. Indeed, it is reasonable to assume that larger departments are characterized, on average, by smaller correlation levels, since they include a higher number of heterogeneous SDSs as well as of ``informal'' research groups. As a consequence, a randomly selected pair of SPs is likely to present a lower correlation compared to an analogous pair from a smaller department. Moreover, in the slightly simplified scenario where each scholar delivers exactly $k$ SPs (in fact, $k$ is a mere multiplier and can be non-integer; see Section~\ref{subsec:anvur}), the share of SP pairs submitted by the same scholar (naturally more inclined to be correlated) is
\[
N_d \cdot {k \choose 2} \bigg/{N_d \choose 2} = \frac{k(k-1)}{N_d-1},
\]
i.e., decreasing with department's size $N_d$.  

The correlation model postulated here is
\begin{equation}\label{eq:corrmodel}
\log\frac{1+\tilde{N}\rho_d}{1-\rho_d} \equiv F(\rho_d) = \alpha + \beta (N_d-1),
\end{equation}
where $\bm{\theta}=(\alpha,\beta)^\top$ is the unknown parameter vector, $F(\rho_d)$ is the resulting linear predictor (unconstrained on the real line), and $\tilde{N}$ denotes the maximum department size. In detail, $F(\rho_d)$ is a modified version of Fisher’s transformation~\citep{Fisher1915}, yielding
\begin{equation}\label{eq:rhotheta}
\rho_d = \frac{\exp\{F(\rho_d)\}-1}{\exp\{F(\rho_d)\}+\tilde{N}}.
\end{equation}
This parameterization guarantees that all correlations lie between $-1/\tilde{N}$ and $1$, ensuring non-negativity of the corresponding standard deviations in~\eqref{eq:vrho}. As a matter of fact, since it is reasonable to assume that $\rho_d$ decreases with $N_d$ (see the discussion above) and the pseudo-Fisher transformation $F(\rho_d)$ is monotonic in $\rho_d$, the expected sign of $\beta$ is negative. This is particularly relevant for index adjustment purposes, as $\beta<0$ ensures that $\rho_d N_d$ tends to a finite limit as $N_d\to\infty$ and, in turn, convergence to Normality of $\tilde{Z}_d$.

Model~\eqref{eq:corrmodel} allows average correlations to differ across departments; it is therefore termed Full Correlation Model (FCM). Its parameter vector $\bm{\theta}$ can be estimated by ML using the observed vector of scaled averages $\bm{\tilde{z}}=(\tilde{z}_1,\dots,\tilde{z}_D)^\top$ based on the log-likelihood
\begin{equation}\label{eq:lk1}
\ell_z(\bm{\tilde{z}};\bm{\theta})=\sum_{d=1}^D -\log\sigma_d+\log\phi\bigl(\tilde{z}_d/\sigma_d\bigr),
\end{equation}
where $\phi(\cdot)$ denotes the standard Normal Probability Density Function (PDF) and $\sigma_d$ depends on $\bm{\theta}$ through~\eqref{eq:vrho} and~\eqref{eq:rhotheta}. Newton–Raphson (NR) or direct numerical maximization can be implemented using the score vector
\[
\bm{s}_z(\bm{\tilde{z}};\bm{\theta}) = \frac{\partial}{\partial\bm{\theta}}\ell_z(\bm{\tilde{z}};\bm{\theta})= \sum_{d=1}^D \begin{pmatrix} 1 \\ N_d-1 \end{pmatrix} \otimes s_{z,d}^{(\alpha)}(\bm{\theta})
\]
and the Hessian matrix
\[
\bm{\mathcal{H}}_z(\bm{\tilde{z}};\bm{\theta}) = \frac{\partial}{\partial\bm{\theta}\partial\bm{\theta}^\top}\ell_z(\bm{\tilde{z}};\bm{\theta})= \sum_{d=1}^D \begin{pmatrix} 1 & N_d-1 \\ N_d-1 & (N_d-1)^2 \end{pmatrix} \otimes h_{z,d}^{(\alpha)}(\bm{\theta}),
\]
where $s_{z,d}^{(\alpha)}(\bm{\theta})$ and $h_{z,d}^{(\alpha)}(\bm{\theta})$ denote the first and second derivatives with respect to $\alpha$ of department $d$’s contribution to~\eqref{eq:lk1}. Their expressions are given by
\[
\begin{split}
s_{z,d}^{(\alpha)}(\bm{\theta}) 
&= \frac{(N_d-1)\delta_d^{(\alpha)}}{2\sigma_d^4}\bigl(\tilde{z}_d^2-\sigma_d^2\bigr)\\[6pt]
h_{z,d}^{(\alpha)}(\bm{\theta}) 
&= s_{z,d}^{(\alpha)}(\bm{\theta})
\Biggl[
\frac{\tilde{N}-\exp\bigl\{\alpha+\beta(N_d-1)\bigr\}}
     {\tilde{N}+\exp\bigl\{\alpha+\beta(N_d-1)\bigr\}}
-\frac{2(N_d-1)\delta_d^{(\alpha)}}{\sigma_d^2}
\Biggr]
-\Biggl[
\frac{(N_d-1)\delta_d^{(\alpha)}}{\sqrt{2}\sigma_d^2}
\Biggr]^2,
\end{split}
\]
where
\begin{equation}\label{eq:dda}
\delta_d^{(\alpha)} 
= \frac{\partial}{\partial \alpha}\rho_d
= \frac{\exp\{F(\rho_d)\}(\tilde{N}+1)}
       {\bigl[\exp\{F(\rho_d)\}+\tilde{N}\bigr]^2}.
\end{equation}
Note that in the above structure derivatives with respect to $\beta$ are always equal to derivatives with respect to $\alpha$ times $N_d-1$.

Once the ML estimate $\hat{\bm{\theta}}$ is available, estimated intra-departmental correlations $\hat{\rho}_d$ can be obtained by replacing $F(\rho_d)$ with $\hat{F}(\rho_d)=\hat{\alpha}+\hat{\beta}(N_d-1)$ in~\eqref{eq:rhotheta}. Also, standard errors can be computed from the above Hessian matrix in order to build confidence intervals and/or test hypotheses on $\alpha$ and $\beta$. In case data provide evidence for $\beta < 0$, it is possible to build the adjusted index
\begin{equation}\label{eq:ispdadj}
\textup{ISPD}_d^{\textsc{fcm}} = \lfloor 200\cdot x_d^{\textsc{fcm}} + 0.5 \rfloor/2 
\end{equation}
where $x_d^{\textsc{fcm}} = \Phi(\tilde{z}_d/\hat{\sigma}_d)$, and $\hat{\sigma}_d = \{1+\hat{\rho}_d(N_d-1)\}^{1/2}$.

Two models can be nested in the FCM as special cases: \emph{i)} the model where average correlations are non-null but identical, and \emph{ii)} the model where all average correlations are null, which is the one implicitly adopted by ANVUR when using the original ISPD. The former, which can be thought of as the counterpart of the above mentioned RIM leading to~\eqref{eq:ispdrim}, is obtained setting $\beta=0$ and is labeled Constant Correlation Model (CCM); the latter is characterized by $\alpha=\beta=0$ and is labeled Null Correlation Model (NCM). Usual Likelihood Ratio Tests (LRTs) can be conducted to assess whether data support the FCM, the CCM or the NCM. With this regard, it is worth to remark that ML estimation of the CCM can be performed by fixing $\beta=0$ in the above quantities, so that the resulting log-likelihood can be maximized with respect to $\alpha$ only. On the contrary, for the NCM maximization is unnecessary, with the model log-likelihood obtained via direct computation of~\eqref{eq:lk1} in $\bm{\theta}=(0,0)^\top$.


\section{Estimation of department-specific variances with public data}\label{sec:estpubl}
As mentioned in Section~\ref{subsec:rho}, for the ANVUR 2017 ranking exercise the ISPDs of all evaluated departments have been published~\citep{Roars}, whereas for the 2022 exercise only the upper tail of the distribution (the top 350 departments) has been released~\citep{MUR2022}. These values represent the only piece of publicly available information: SP-level scores (micro-data) have not been disclosed for privacy reasons, and ISPDs cannot be exactly back-transformed into scaled averages because of the rounding step embedded in the index construction. Consequently, the ML procedure of Section~\ref{sec:correction}  - which relies on scaled averages - has to be intended as a correction tool for complete data holders (like ANVUR officers) rather than a method to estimate and test the FCM based on publicly available data only. The latter, however, remains essential to quantify over-dispersion of VD scaled averages and, in turn, the unfairness level of the original ISPD. Therefore, it is developed in this section, still in a ML framework.

The key step for the above task consists in deriving the distribution of the transformation $X_d=\Phi(\tilde{Z}_d)$, where the random VD scaled average $\tilde{Z}_d$ follows (approximately) a centered Normal distribution with variance $\sigma^2_d$. In what follows, we introduce this novel probability distribution and its left-truncated version (Section~\ref{subsec:betoidal}), which form the basis for ML estimation of the FCM with 2017 (Section~\ref{subsec:coarse}) and 2022 (Section~\ref{subsec:coarsetrunc}) data. Estimation results are reported in Section~\ref{subsec:res}.


\subsection{The Betoidal distribution}\label{subsec:betoidal}
Consider $\tilde{Z}\sim N(0,\sigma^2)$, at least approximately, and $X=\Phi(\tilde{Z})$. By standard results on transformations of random varaibles, the PDF of $X$ for $x\in(0,1)$ is 
\begin{equation}\label{eq:pdf}
\begin{split}
f_X(x;\sigma) &= \frac{\sqrt{2\pi}}{\sigma}\phi\Biggl(\frac{\Phi^{-1}(x)}{\sigma} \Biggr)\exp\{[\tu{erf}^{-1}(2x-1)]^2\} \\
&= \phi\Bigl(\Phi^{-1}(x)/\sigma \Bigr)\cdot\Bigl\{\sigma\cdot \phi\Bigl(\Phi^{-1}(x) \Bigr)\Bigr\}^{-1},
\end{split}
\end{equation}
where $\tu{erf}^{-1}(\cdot)$ is the inverse of the error function
\[
\tu{erf}(x) = \frac{2}{\sqrt{\pi}}\int_0^x e^{-t^2}dt.
\]
The corresponding CDF and quantile function are, respectively,
\begin{equation}\label{eq:cdf}
\begin{split}
F_X(x;\sigma) &= P(X\leq x) = P(\Phi(Z) \leq x) = P(Z \leq \Phi^{-1}(x)) = F_Z(\Phi^{-1}(x)) \\
&= F_Z(\sqrt{2}\cdot\tu{erf}^{-1}(2x-1)) = \frac{1}{2}\Biggl\{1+\tu{erf}\Biggl(\frac{\tu{erf}^{-1}(2x-1)}{\sigma} \Biggr) \Biggr\},
\end{split}
\end{equation}
and, for $q\in [0,1]$,
\begin{equation}\label{eq:qfunctbet}
F_X^{-1}(q;\sigma) = \frac{1+\tu{erf}\{\sigma\cdot\tu{erf}^{-1}(2q-1)\}}{2}.
\end{equation}

\begin{figure}[tb]
\centering{
\subfloat[][\label{fig:bet}]{\includegraphics[scale=0.53]{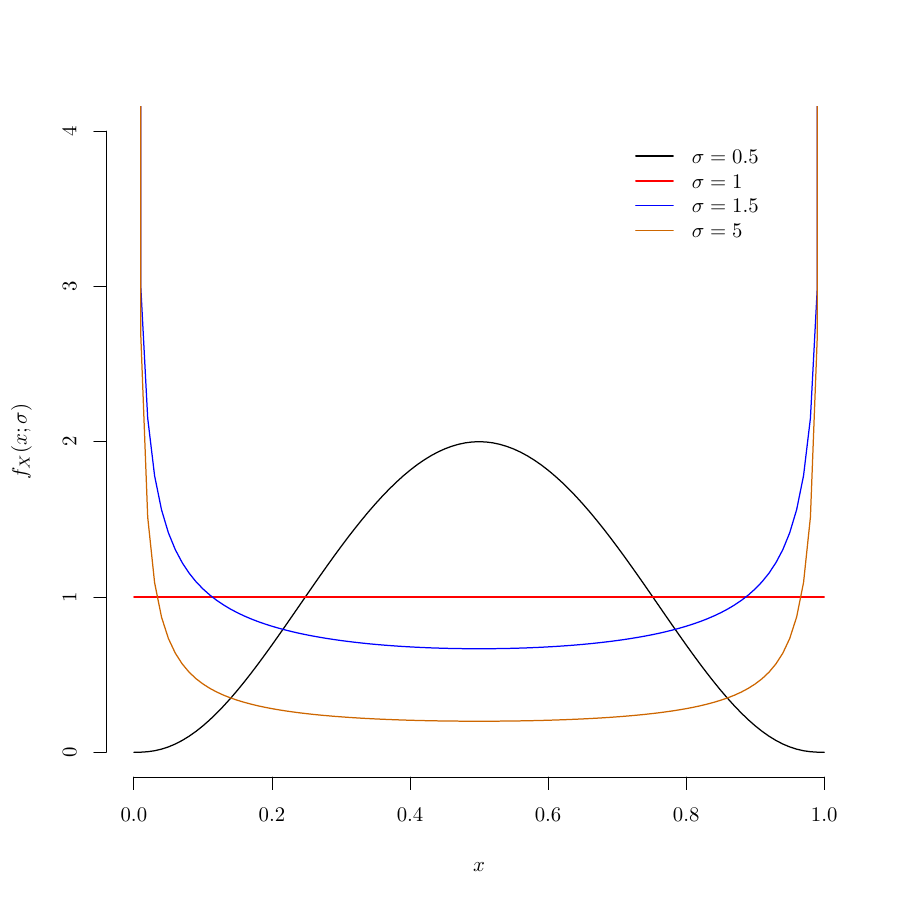}}\quad
\subfloat[][\label{fig:sa}]{\includegraphics[scale=0.53]{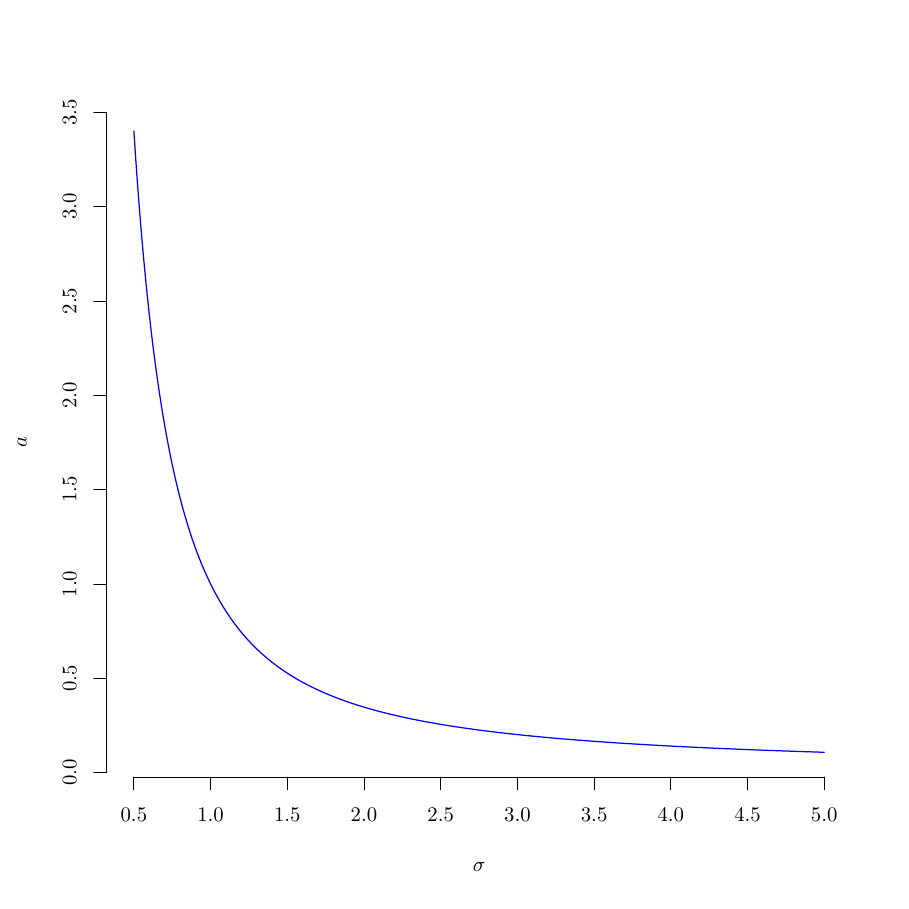}}
}
\caption{(a) Betoidal density for various levels of $\sigma$, and (b) relation between $\sigma$ and $a$ for two random variables $X\sim \textup{Betoidal}(\sigma)$ and $Y\sim\textup{Beta}(a,a)$ with $V(X)=V(Y)$.}
\label{fig:betsa}
\end{figure}

When $\sigma=1$, this distribution reduces to the Uniform$(0,1)$, since $f_X(x;1)=1$, $F_X(x;1)=x$, and $F_X^{-1}(q;1)=q$. 
For $\sigma\neq 1$, the density becomes U-shaped when $\sigma > 1$ and bell-shaped when $\sigma < 1$.
The shape resembles that of a symmetric $\tu{Beta}(a,a)$ distribution with $a< 1$ and $a>1$, respectively  (see Figure~\ref{fig:bet}), though it is not identical. 
Because of this similarity, we refer to the distribution, indexed by $\sigma$, as to the {\em Betoidal} distribution and write $X\sim\textup{Betoidal}(\sigma)$.

\begin{figure}[tb]
\centering{
\subfloat[][\label{fig:betabet05}]{\includegraphics[scale=0.53]{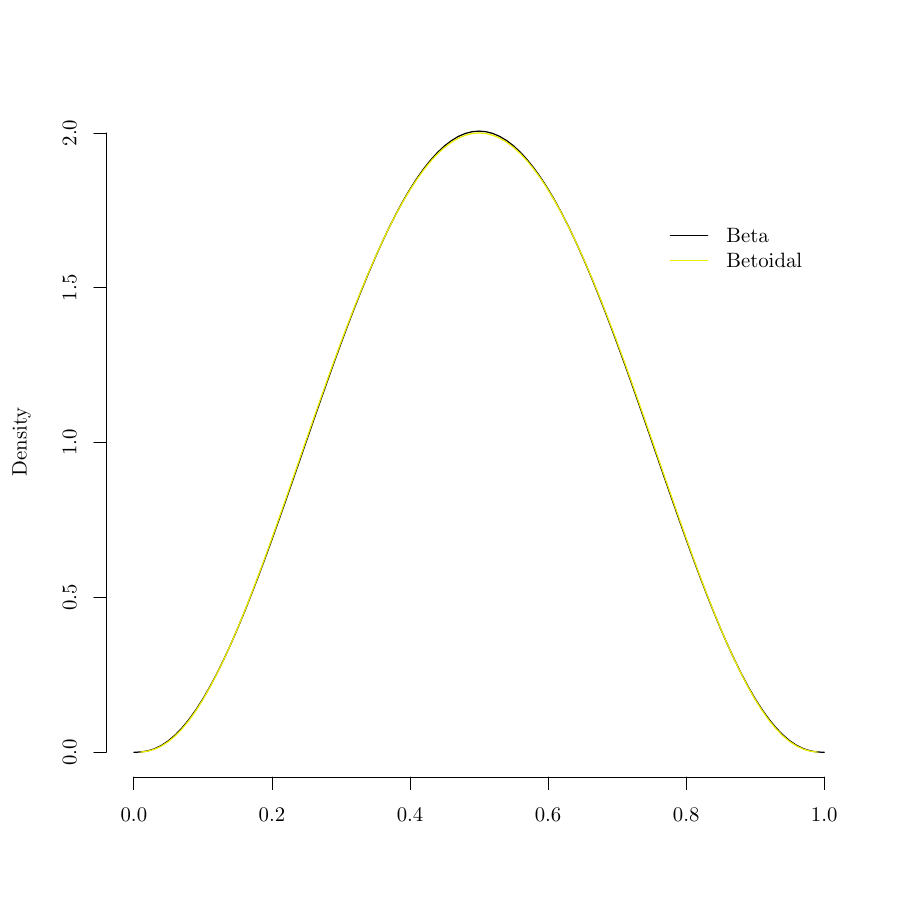}}\quad
\subfloat[][\label{fig:betabet25}]{\includegraphics[scale=0.53]{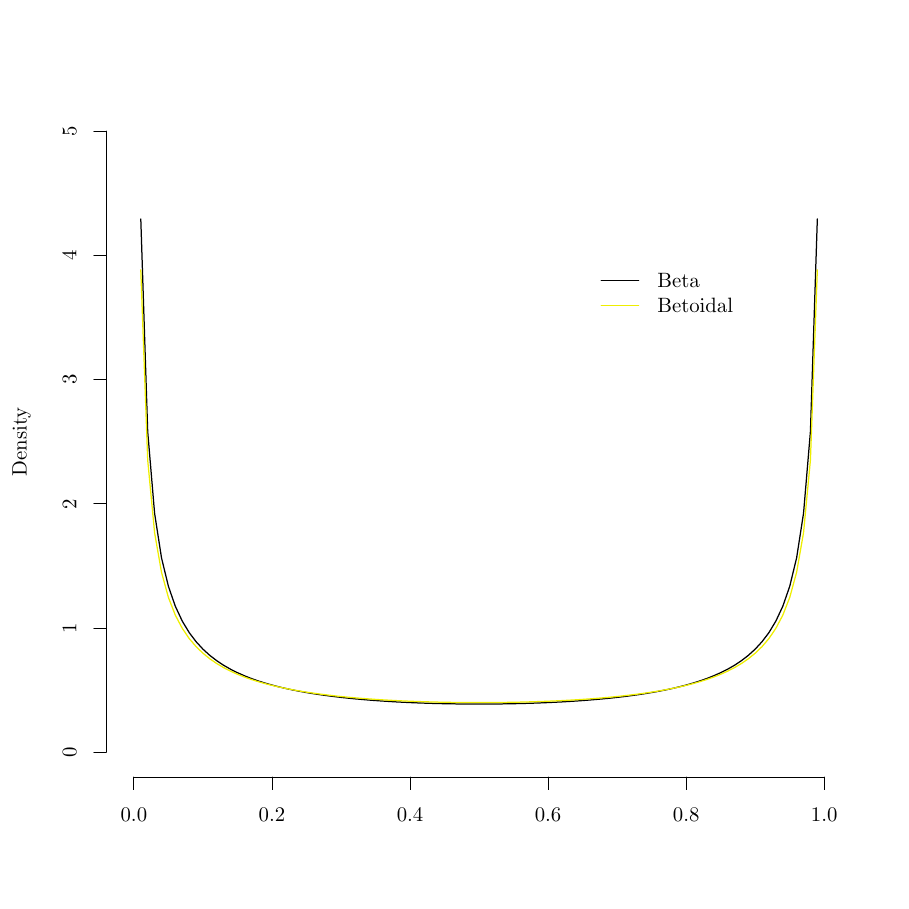}}
}
\caption{Density of $X\sim \textup{Betoidal}(\sigma)$ and $Y\sim\textup{Beta}(a,a)$ with $V(X)=V(Y)$ and (a) $\sigma=0.5$ ($a=3.4005$), and (b) $\sigma=2.5$ ($a=0.2568$).}
\label{fig:betabet}
\end{figure}

Both the expected value and the median of $X$ equal $0.5$, while the variance is
\begin{equation}\label{eq:varbetoidal}
V(X)=\frac{1-8\cdot T(0,1/\sqrt{2\sigma^2+1})}{4}=\frac{1}{\pi}\arctan(\sqrt{1+2\sigma^2})-\frac{1}{4},
\end{equation}
where $T(h,c)$ denotes Owen’s $T$ function~\citep{Owen1956}; see Appendix~\ref{app:moments} for derivations and further details on ML estimation of $\sigma$. Equating~\eqref{eq:varbetoidal} to the variance of a symmetric Beta distribution, given by $[4\cdot(2a+1)]^{-1}$, yields
\[
a=\frac{\{1-8\cdot T(0,1/\sqrt{2\sigma^2+1})\}^{-1}-1}{2}.
\]
The relationship between $\sigma$ and $a$ is illustrated in Figure~\ref{fig:sa}. With equal variance, the Beta and Betoidal densities are nearly indistinguishable (Figure~\ref{fig:betabet}), although small discrepancies arise near the center and at the boundaries.  

When only the upper tail of the ISPD values is observed, as in the 2022 case, it is necessary to consider a left-truncated version of the distribution. To this end, we indicate by $X^{\star}\sim\textup{LT-Betoidal}(\sigma,x^{\star})$ a Betoidal random variable truncated below a given value $x^{\star}$. Its support is $[x^{\star},1)$, and its PDF and CDF are, respectively,
\[
f_{X^{\star}}(x;\sigma,x^{\star})=\frac{f_X(x;\sigma)}{1-F_X(x^{\star};\sigma)}
\]
and
\begin{equation}\label{eq:cdftrunc}
F_{X^{\star}}(x;\sigma,x^{\star})=\frac{F_X(x;\sigma)-F_X(x^{\star};\sigma)}{1-F_X(x^{\star};\sigma)}.
\end{equation}


\subsection{ML estimation for 2017 data}\label{subsec:coarse}
Because of the rounding mechanism discussed earlier in the paper, ISPDs take values in the grid
\[
\mathcal{S}=\{s_1,s_2,s_3,\dots,s_{J-2},s_{J-1},s_J\}=\{0, 0.5, 1,\dots,99, 99.5, 100\}, 
\]
where $J=201$ (see Section~\ref{subsec:rho}). Thus, they can be treated as realizations from the random vector $\bm{I}=(I_1,\dots,I_D)^\top$, where each component $I_d$ has discrete support $\mathcal{S}$. Since the event $[I_d=s_j]$ corresponds to the event $[\tilde{s}_{j-}< X_d \leq \tilde{s}_{j+}]$, where $X_d\sim\text{Betoidal}(\sigma_d)$ and
\begin{equation}\label{eq:stilde}
\begin{split}
\tilde{s}_{j+} &= (s_j+0.25)/100 \\
\tilde{s}_{j-} &= (s_j-0.25)/100,
\end{split}
\end{equation}
the probability $P(I_d=s_j)$ can be linked to the FCM parameter vector $\bm{\theta}$ through the Betoidal CDF in~\eqref{eq:cdf}. In detail, we have
\begin{equation}\label{eq:pd}
\begin{split}
\pi_{dj}(\bm{\theta}) &\equiv P(I_d=s_j)=F_{X}(\tilde{s}_{j+};\sigma_d)-F_{X}(\tilde{s}_{j-};\sigma_d) \\
&=\frac{1}{2}\Biggl[\textup{erf}\biggl(\frac{e(\tilde{s}_{j+})}{\sigma_d}\biggr)-\textup{erf}\biggl(\frac{e(\tilde{s}_{j-})}{\sigma_d} \biggr)\Biggr],
\end{split}
\end{equation}
where $e(x)$ is a shorthand for $\textup{erf}^{-1}(2x-1)$ and, like in Section~\ref{sec:correction}, dependence on $\bm{\theta}$ is conveyed by $\sigma_d$ via~\eqref{eq:vrho} and~\eqref{eq:rhotheta}. In this setting, the likelihood of the observed data $\bm{\iota}=(\text{ISPD}_1,\dots,\text{ISPD}_D)^\top$ is
\[
P(\bm{I}=\bm{\iota};\bm{\theta}) =\prod_{d=1}^D\prod_{j=1}^J \pi_{dj}(\bm{\theta})^{\mathbb{I}(\text{ISPD}_d=s_j)},
\]
where $\mathbb{I}(\cdot)$ denotes the indicator function. The corresponding log-likelihood is
\[
\ell_{\iota}(\bm{\iota};\bm{\theta}) =\sum_{d=1}^D\sum_{j=1}^J \mathbb{I}(\text{ISPD}_d=s_j)\log \pi_{dj}(\bm{\theta}).
\]
The associated score vector and Hessian matrix are given by
\[
\bm{s}_{\iota}(\bm{\iota};\bm{\theta})=\sum_{d=1}^D\sum_{j=1}^J\mathbb{I}(\text{ISPD}_d=s_j)\begin{pmatrix} 1 \\ N_d-1 \end{pmatrix} \otimes s_{\iota,dj}^{(\alpha)}(\bm{\theta})
\]
and
\[
\bm{\mathcal{H}}_{\iota}(\bm{\iota};\bm{\theta}) = \sum_{d=1}^D\sum_{j=1}^J\mathbb{I}(\text{ISPD}=s_j) \begin{pmatrix} 1 & N_d-1 \\ N_d-1 & (N_d-1)^2 \end{pmatrix} \otimes h_{\iota,dj}^{(\alpha)}(\bm{\theta}),
\]
with the explicit expressions of
\[
s_{\iota,dj}^{(\alpha)}(\bm{\theta}) = \frac{\partial}{\partial\alpha} \log \pi_{dj}(\bm{\theta})  \qquad\qquad\text{and}\qquad\qquad h_{\iota,dj}^{(\alpha)}(\bm{\theta}) = \frac{\partial^2}{\partial^2\alpha} \log \pi_{dj}(\bm{\theta}),
\]
reported in Appendix~\ref{app:nr}. Again, ML estimation of $\bm{\theta}$ can be carried out either by direct or NR optimization. With regard to inference and model testing (FCM versus CCM and/or NCM), the same considerations of Section~\ref{sec:correction} apply.


\subsection{ML estimation for 2022 data}\label{subsec:coarsetrunc}

In the 2022 ranking, the rounding mechanism is the same as in the 2017 ranking, but truncation occurs since only the highest $D^{\star}=350$ ISPD values (namely, those greater or equal than 73) are available. In practice, the ISPD vector $\bm{\iota}^{\star}=(\text{ISPD}_1,\dots,\text{ISPD}_{D^{\star}})^\top$ can be treated as a draw from $\bm{I}^{\star}=(I^{\star}_1,\dots,I^{\star}_{D^{\star}})^\top$, with the support of each random variable $I^{\star}_d$ reduced to
\[
\mathcal{S}^{\star}=\{s_{J^{\star}},s_{J^{\star}+1},s_{J^{\star}+2},\dots,s_{J-2},s_{J-1},s_{J}\}=\{73, 73.5, 74,\dots,99, 99.5, 100\} 
\]
($J^{\star}=146$). In parallel with Section~\ref{subsec:coarse} it is possible to equate the event $[I^{\star}_d=s_j]$ to the event $[\tilde{s}_{j-} < X^{\star}_d \leq \tilde{s}_{j+}]$, where $j=J^{\star},\dots,J$, $\tilde{s}_{j-}$ and $\tilde{s}_{j+}$ are as in~\eqref{eq:stilde} and $X^{\star}_d\sim\text{LT-Betoidal}(\sigma_d,\tilde{s}_{J^{\star}-})$. In the truncated distribution, the truncation point is set to $\tilde{s}_{J^{\star}-}=0.7275$ since ISPDs are always rounded to the closest semi-integer; see~\eqref{GEM1}. Thus, the link between $P(I^{\star}_d=s_j)$ and the FCM vector $\bm{\theta}$ now operates through the CDF in~\eqref{eq:cdftrunc}, that is,
\begin{equation}\label{eq:pstard}
\begin{split}
\pi^{\star}_{dj}(\bm{\theta},\tilde{s}_{J^{\star}-}) &\equiv P(I^{\star}_d=s_j) =F_{X^{\star}}(\tilde{s}_{j+};\sigma_d,\tilde{s}_{J^{\star}-})-F_{X^{\star}}(\tilde{s}_{j-};\sigma_d,\tilde{s}_{J^{\star}-}), \\
&= \pi_{dj}(\bm{\theta})/\{1-F_X(\tilde{s}_{J^{\star}-};\sigma_d)\}.
\end{split}
\end{equation}
As usual, dependence on $\bm{\theta}$ operates through $\sigma_d$. ML estimation follows the same pattern as in Section~\ref{subsec:coarse}. In detail, the likelihood function is
\[
P(\bm{I}^{\star}=\bm{\iota}^{\star};\bm{\theta},\tilde{s}_{J^{\star}-})=\prod_{d=1}^{D^{\star}}\prod_{j=J^{\star}}^J \pi^{\star}_{dj}(\bm{\theta},\tilde{s}_{J^{\star}-})^{\mathbb{I}(\text{ISPD}_d=s_j)},
\]
the corresponding log-likelihood is
\[
\ell_{\iota^{\star}}(\bm{\iota}^{\star};\bm{\theta},\tilde{s}_{J^{\star}-})=\sum_{d=1}^{D^{\star}}\sum_{j=J^{\star}}^J \mathbb{I}(\text{ISPD}_d=s_j)\log \pi^{\star}_{dj}(\bm{\theta},\tilde{s}_{J^{\star}-})
\]
and the associated score vector and Hessian matrix are respectively given by
\[
\bm{s}_{\iota^{\star}}(\bm{\iota}^{\star};\bm{\theta},\tilde{s}_{J^{\star}-})=\sum_{d=1}^{D^{\star}}\sum_{j=J^{\star}}^J\mathbb{I}(\text{ISPD}_d=s_j)\begin{pmatrix} 1 \\ N_d-1 \end{pmatrix} \otimes s_{\iota^{\star},dj}^{(\alpha)}(\bm{\theta},\tilde{s}_{J^{\star}-})
\]
and
\[
\bm{\mathcal{H}}_{\iota^{\star}}(\bm{\iota}^{\star};\bm{\theta},x^{\star}) = \sum_{d=1}^{D^{\star}}\sum_{j=J^{\star}}^J\mathbb{I}(\text{ISPD}_d=s_j) \begin{pmatrix} 1 & N_d-1 \\ N_d-1 & (N_d-1)^2 \end{pmatrix} \otimes h_{\iota^{\star},dj}^{(\alpha)}(\bm{\theta},\tilde{s}_{J^{\star}-}),
\]
where the expressions of
\[
s_{\iota^{\star},dj}^{(\alpha)}(\bm{\theta},\tilde{s}_{J^{\star}-}) = \frac{\partial}{\partial\alpha} \log \pi^{\star}_{dj}(\bm{\theta},\tilde{s}_{J^{\star}-})  \qquad\qquad\text{and}\qquad\qquad h_{\iota^{\star},dj}^{(\alpha)}(\bm{\theta},\tilde{s}_{J^{\star}-}) = \frac{\partial^2}{\partial^2\alpha} \log \pi^{\star}_{dj}(\bm{\theta},\tilde{s}_{J^{\star}-}),
\]
are in Appendix~\ref{app:nr}. Parameter estimation, inference and model selection involve standard ML theory, for which we refer to Sections~\ref{sec:correction} and~\ref{subsec:coarse}.

\subsection{Results from the ISPD data of ANVUR 2017 and 2022 rankings}\label{subsec:res}

Table~\ref{tab:res} reports the ML estimation results obtained from the available ISPDs for the ANVUR 2017 and 2022 rankings, based on the models described in Sections~\ref{subsec:coarse} and~\ref{subsec:coarsetrunc}, respectively. The left panel refers to the 2017 data, while the right panel to the 2022 data. For both panels, the first row summarizes the distribution of the number of submitted SPs, $N_d$, across departments. The central rows report results for the FCMs as in~\eqref{eq:corrmodel}. For these models, the ML estimates of $\alpha$ and $\beta$ are presented together with their standard errors and $p$-values. In addition, summary statistics of the estimated intra-departmental correlations ($\hat{\rho}_d$) and standard deviations ($\hat{\sigma}_d$) are provided. The bottom rows display results for the NCMs ($\alpha=0=\beta$) and the CCMs ($\beta=0$). For both specifications, LRT statistics comparing each model with the corresponding FCM, along with their $p$-values, are reported. For CCMs, the estimates of $\alpha$ and their standard errors are also shown. For each model, direct maximization of the log-likelihood was performed via the Broyden-Fletcher-Goldfarb-Shanno (BFGS) algorithm~\citep{Fletcher1987book}. NR optimization yielded virtually identical results. In both cases, a multiple starting-point strategy was adopted to ensure adequate exploration of the parameter space and to mitigate potential numerical issues or convergence to local maxima.

\begin{table}[tb]
\centering
\begin{tabularx}{\textwidth}{YYYcYYYYYcYY}
    \toprule
\multicolumn{6}{c}{{\bf 2017 data} (766 departments)} & \multicolumn{6}{c}{{\bf 2022 data} (350 departments)} \\
\cmidrule(lr){1-6}
\cmidrule(lr){7-12}
        \multicolumn{6}{c}{$N_d$} &  \multicolumn{6}{c}{$N_d$} \\ 
        min & Q1 & Q2 & mean & Q3 & max & min & Q1 & Q2 & mean & Q3 & max \\ 
        24 & 96 & 120 & 130.6 & 153.5 & 464 & 78 & 159 & 198 & 219.8 & 254.2 & 615 \\ 
\cmidrule(lr){1-6}
\cmidrule(lr){7-12}
        \multicolumn{6}{c}{{\bf FCM} (log-lk = -3257.902)} & \multicolumn{6}{c}{{\bf FCM} (log-lk = -950.0124)} \\
\cmidrule(lr){1-6}
\cmidrule(lr){7-12}
        $\hat{\alpha}$ & s.e. & $p$ & $\hat{\beta}$ & s.e. & $p$ & $\hat{\alpha}$ & s.e. & $p$ & $\hat{\beta}$ & s.e. & $p$ \\ 
        3.7527 & 0.2043 & 0.0000 & -0.0038 & 0.0014 & 0.0086 & 3.6793 & 0.2828 & 0.0000 & -0.0023 & 0.0012 & 0.0458 \\
\cmidrule(lr){1-6}
\cmidrule(lr){7-12}
     \multicolumn{6}{c}{$\hat{\rho}_d$} & \multicolumn{6}{c}{$\hat{\rho}_d$} \\ 
        min & Q1 & Q2 & mean & Q3 & max & min & Q1 & Q2 & mean & Q3 & max \\ 
        0.0138 & 0.0473 & 0.0536 & 0.0522 & 0.0585 & 0.0759 & 0.0135 & 0.0329 & 0.0376 & 0.0363 & 0.0412 & 0.0496 \\
\cmidrule(lr){1-6}
\cmidrule(lr){7-12}
        \multicolumn{6}{c}{$\hat{\sigma}_d$} &  \multicolumn{6}{c}{$\hat{\sigma}_d$} \\ 
           min & Q1            & Q2             & mean & Q3 & max & min & Q1 & Q2 & mean & Q3 & max \\ 
1.657 & 2.561 & 2.715 & 2.705 & 2.860 & 3.027 & 2.195 & 2.739 & 2.899 & 2.888 & 3.052 & 3.202 \\ 
\cmidrule(lr){1-6}
\cmidrule(lr){7-12}
        \multicolumn{2}{c}{{\bf NCM}} & \multicolumn{4}{c}{{\bf CCM}} & \multicolumn{2}{c}{{\bf NCM}} & \multicolumn{4}{c}{{\bf CCM}} \\ 
\cmidrule(lr){1-2}
\cmidrule(lr){3-6}
\cmidrule(lr){7-8}
\cmidrule(lr){9-12}
        $\chi^2_{\text{{\tiny LRT}}}$ & $p$ & $\hat{\alpha}$ & s.e. & $\chi^2_{\text{{\tiny LRT}}}$ & $p$ & $\chi^2_{\text{{\tiny LRT}}}$ & $p$ & $\hat{\alpha}$ & s.e. & $\chi^2_{\text{{\tiny LRT}}}$ & $p$ \\ 
       1931.2 & 0.000 & 3.2578 & 0.0728 & 6.6001 & 0.0102 & 1095.2 & 0.000 & 3.1605 & 0.1010 & 3.7571 & 0.0526 \\
\bottomrule
    \end{tabularx}
    \caption{Estimation results for the ANVUR 2017 and 2022 ranking exercises. FCM: Full Correlation Model, CCM: Constant Correlation Model, NCM: Null Correlation Model, $\chi^2_{\text{{\tiny LRT}}}$: Likelihood Ratio Test statistic, s.e.: standard error, $p$: p-value, Q1: first quartile, Q2: median, Q3: third quartile.}\label{tab:res}
\end{table}

In both FCMs, the estimate of $\beta$ is negative. LRTs comparing NCMs to FCMs provide overwhelming evidence against the null hypothesis of no correlation (with $\chi^2_{\text{{\tiny LRT}}}$ statistics exceeding 1000 and $p$-values effectively equal to zero), thereby confirming the presence of intra-departmental dependence in both datasets. Although at different significance levels, the data also tend to favor FCMs over CCMs (where the estimated common correlation is 0.0511 in 2017 and 0.0354 in 2022): the LRT statistic equals 6.6001 ($p$-value = 0.0102) in the 2017 exercise and 3.7571 ($p$-value = 0.0526) in the 2022 exercise. These findings are consistent with the Wald tests on $\beta$ within the FCMs, whose $p$-values are equal to 0.0086 and 0.0458 ($H_1:\,\beta\neq0$) or to 0.0043 and 0.0229 ($H_1:\,\beta < 0$). These pieces of evidence in favor of $\beta < 0$ support the key hypothesis that intra-departmental average correlation decreases as department size increases; see the related discussion in Section~\ref{sec:correction}.

In VQR 2015-2019 (generating 2022 ISPD data), the number of submitted products is substantially larger than in VQR 2010-2014 (used for the 2017 ranking). This increase reflects both the change in the VQR multiplier $k$ (from 2 to 3; see Section~\ref{subsec:anvur}) and the effect of an Italian legislative decree enacted in 2014, which set the minimum department size at 35 units. As a consequence, the intra-departmental correlations estimated from the 2022 ranking are lower on average, although this reduction is partly offset by the smaller absolute magnitude of $\hat{\beta}$. Conversely, the estimated standard deviations $\hat{\sigma}_d$ are slightly larger in 2022, since the increase in department sizes $N_d$ more than compensates the decline in the estimated correlations $\hat{\rho}_d$; see Equation~\eqref{eq:vrho}. These results are consistent with the patterns observed in Figures~\ref{fig:ispd17} and~\ref{fig:ispd22}, and confirm the greater polarization of ISPDs in the 2022 ranking compared to the 2017 ranking (see Section~\ref{subsec:rho}).


\section{Simulation study}\label{sec:simulazio}

This section presents a simulation study designed to assess the effectiveness of the ISPD adjustments introduced in Section~\ref{sec:correction}, in a setting closely resembling the ANVUR 2017 exercise. To this end, we consider the same number of departments ($D=766$) and replicate the empirical of $N_d$ ($d=1,\dots,D$). For simplicity, the SDS classification (see Section~\ref{subsec:anvur}) is ignored, and SP standardized scores are generated through random draws from a unique standardized distribution with support $(-1.69580, -1.06773, -0.12561, 0.81650, 1.44457)$ and probability vector $(0.1, 0.2, 0.3, 0.25, 0.15)$, corresponding to the marginal distribution of standardized scores observed in VQR 2010-2014. 

For each department $d$, $N_d$ standardized scores are generated so as to attain a pre-specified intra-departmental correlation level $\rho_d$; see Appendix~\ref{app:simsdgp} for details on the data-generating process. The value of $\rho_d$ is fixed according to four scenarios. In the baseline scenario, $\rho_d$ follows Equation~\eqref{eq:rhotheta}, with parameter vector $\bm{\theta}=(\alpha,\beta)^\top$ set equal to $(3.752,-0.00376)^\top$, i.e., the ML estimate obtained from the empirical analysis in Section~\ref{subsec:res}. To mimic potential deviations from the theoretical size–correlation relationship in~\eqref{eq:rhotheta}, three additional scenarios introduce multiplicative random perturbations of each $\rho_d$ drawn from: 
\emph{i}) a $U(0.9,1.1)$ distribution (say small perturbation), 
\emph{ii}) a $U(0.75,1.25)$ distribution (say medium perturbation), and 
\emph{iii}) a $U(0.5,1.5)$ distribution (say large perturbation).

For each scenario, 1000 datasets are generated. Within each dataset, $N_d$ standardized scores are assigned to department $d$ according to the fixed intra-departmental correlation $\rho_d$. The resulting scaled average $\tilde{z}_d$ is then computed, along with the following indices:
\begin{itemize}
\item $\text{ISPD}^{\textsc{theo}}$ in~\eqref{eq:ispdtheo};
\item $\text{ISPD}$ in~\eqref{GEM1}, that is, the original unadjusted ANVUR index;
\item $\text{ISPD}^{\textsc{np}}$ in~\eqref{eq:ispdmicro}, computed after forcing $\hat{\rho}_d^{\textsc{np}}$ to lie in $[0,1]$ when necessary;
\item $\text{ISPD}^{\textsc{fcm}}$ in~\eqref{eq:ispdadj};
\item $\text{ISPD}^{\textsc{rim}}$ in~\eqref{eq:ispdrim}, which \emph{de facto} corresponds to assuming a CCM (see Section~\ref{sec:correction}).
\end{itemize}
$\text{ISPD}^{\textsc{theo}}$ is computed using the theoretical standard deviations $\sigma_d$ and therefore represents the infeasible benchmark. The performance of the other indices is evaluated relative to it, using two metrics defined on the 0–100 scale: the Mean Absolute Deviation (MAD) and the Percentage of Discordant Comparisons (PDC). Denoting by $\text{ISPD}^{*}$ the index under evaluation, the MAD is defined as
\[
\text{MAD} = \frac{1}{D}\sum_{d=1}^D 
\left| \text{ISPD}_d^* - \text{ISPD}_d^{\text{tr}} \right|,
\]
thereby representing the average absolute deviation from the benchmark across departments. On the other hand, the PDC is defined as
\[
\text{PDC} = \frac{200}{D(D-1)}\sum_{d=2}^D\sum_{d'=1}^{d-1} 
\mathbb{I}\left\{
\text{sgn}(\text{ISPD}_d^*-\text{ISPD}_{d'}^*)
\neq 
\text{sgn}(\text{ISPD}_d^{\text{tr}}-\text{ISPD}_{d'}^{\text{tr}})
\right\}
\]
(with $\text{sgn}(\cdot)$ denoting the sign function); it measures the percentage of department pairs for which the ordering induced by $\text{ISPD}^*$ differs from the benchmark ordering.

\begin{table}[tb]
    \centering
    \begin{tabularx}{\textwidth}{cYYYYYYYYYYYY}
\toprule
\multicolumn{13}{l}{{\bf Mean Absolute Deviation (MAD)}} \\[5pt]
	& \multicolumn{6}{c}{{\bf Null perturbation}} & \multicolumn{6}{c}{{\bf Small perturbation}} \\
	\cmidrule(lr){2-7}
	\cmidrule(lr){8-13}
         & min & Q1 & Q2 & mean & Q3 & max & min & Q1 & Q2 & mean & Q3 & max \\ 
        $\text{ISPD}$ & 12.88 & 13.49 & 13.67 & 13.66 & 13.84 & 14.63 & 12.83 & 13.49 & 13.66 & 13.66 & 13.82 & 14.44 \\ 
        $\text{ISPD}^{\textsc{np}}$ & 8.83 & 9.35 & 9.49 & 9.49 & 9.63 & 10.08 & 8.85 & 9.36 & 9.49 & 9.49 & 9.63 & 10.05 \\ 
      	$\text{ISPD}^{\textsc{rim}}$ & 0.85 & 0.93 & 0.97 & 1.00 & 1.04 & 1.68 & 0.92 & 0.99 & 1.02 & 1.05 & 1.08 & 1.63 \\ 
   $\text{ISPD}^{\textsc{fcm}}$ & 0.02 & 0.26 & 0.39 & 0.43 & 0.58 & 1.49 & 0.31 & 0.41 & 0.50 & 0.54 & 0.62 & 1.33 \\
		\cmidrule(lr){2-7}
	\cmidrule(lr){8-13}
	& \multicolumn{6}{c}{{\bf Medium perturbation}} & \multicolumn{6}{c}{{\bf Large perturbation}} \\
		\cmidrule(lr){2-7}
	\cmidrule(lr){8-13}
        ~ & min & Q1 & Q2 & mean & Q3 & max & min & Q1 & Q2 & mean & Q3 & max \\ 
        $\text{ISPD}$ & 12.80 & 13.44 & 13.62 & 13.62 & 13.79 & 14.37 & 12.75 & 13.30 & 13.47 & 13.47 & 13.64 & 14.14 \\ 
        $\text{ISPD}^{\textsc{np}}$ & 8.88 & 9.32 & 9.46 & 9.46 & 9.59 & 10.01 & 8.73 & 9.26 & 9.38 & 9.39 & 9.52 & 10.09 \\ 
      	$\text{ISPD}^{\textsc{rim}}$ & 1.16 & 1.27 & 1.29 & 1.31 & 1.33 & 1.86 & 1.84 & 1.99 & 2.02 & 2.02 & 2.05 & 2.27 \\
  $\text{ISPD}^{\textsc{fcm}}$ & 0.80 & 0.88 & 0.92 & 0.94 & 0.98 & 1.59 & 1.60 & 1.73 & 1.77 & 1.78 & 1.81 & 2.12 \\ 
        \midrule
       \multicolumn{13}{l}{{\bf Percentage of Discordant Comparisons (PDC)}} \\ [5pt]
        	& \multicolumn{6}{c}{{\bf Null perturbation}} & \multicolumn{6}{c}{{\bf Small perturbation}} \\
	\cmidrule(lr){2-7}
	\cmidrule(lr){8-13}
        & min & Q1 & Q2 & mean & Q3 & max & min & Q1 & Q2 & mean & Q3 & max \\ 
        $\text{ISPD}$ & 4.62 & 5.46 & 5.76 & 5.76 & 6.03 & 7.14 & 4.69 & 5.52 & 5.78 & 5.81 & 6.11 & 7.23 \\ 
        $\text{ISPD}^{\textsc{np}}$ & 13.54 & 15.64 & 16.2 & 16.18 & 16.71 & 19.1 & 14.14 & 15.68 & 16.24 & 16.27 & 16.84 & 19.36 \\ 
       $\text{ISPD}^{\textsc{rim}}$ & 1.51 & 1.68 & 1.73 & 1.73 & 1.77 & 1.95 & 1.59 & 1.79 & 1.83 & 1.84 & 1.88 & 2.09 \\ 
  $\text{ISPD}^{\textsc{fcm}}$ & 0.08 & 0.43 & 0.60 & 0.65 & 0.80 & 1.79 & 0.72 & 0.84 & 0.91 & 0.97 & 1.07 & 1.95 \\ 
        	\cmidrule(lr){2-7}
	\cmidrule(lr){8-13}
		& \multicolumn{6}{c}{{\bf Medium perturbation}} & \multicolumn{6}{c}{{\bf Large perturbation}} \\
		\cmidrule(lr){2-7}
	\cmidrule(lr){8-13}
	~ & min & Q1 & Q2 & mean & Q3 & max & min & Q1 & Q2 & mean & Q3 & max \\ 
        $\text{ISPD}$ & 4.67 & 5.74 & 6.03 & 6.03 & 6.31 & 7.72 & 5.51 & 6.33 & 6.58 & 6.61 & 6.87 & 7.92 \\ 
        $\text{ISPD}^{\textsc{np}}$ & 13.73 & 15.73 & 16.24 & 16.26 & 16.76 & 19.24 & 13.97 & 15.64 & 16.16 & 16.19 & 16.72 & 18.97 \\ 
       $\text{ISPD}^{\textsc{rim}}$ & 2.02 & 2.19 & 2.24 & 2.24 & 2.29 & 2.50 & 2.87 & 3.13 & 3.20 & 3.20 & 3.27 & 3.56 \\  
  $\text{ISPD}^{\textsc{fcm}}$ & 1.43 & 1.57 & 1.62 & 1.64 & 1.69 & 2.32 & 2.51 & 2.75 & 2.81 & 2.82 & 2.89 & 3.47 \\ 
        \bottomrule
    \end{tabularx}
        \caption{Simulation results for the original $\text{ISPD}$ and the three adjusted indices $\text{ISPD}^{\textsc{np}}$, $\text{ISPD}^{\textsc{rim}}$ and $\text{ISPD}^{\textsc{fcm}}$. Summary of the distribution (across the 1000 replications) of the Mean Absolute Deviation (MAD) and the Percentage of Discordant Comparisons (PDC) for the four perturbation levels (null, small, medium and large). Q1: first quartile, Q2: median, Q3: third quartile.}\label{tab:sims}
\end{table}


\begin{figure}[tb]
\centering{
\subfloat[][\label{fig:histispd_original}]{\includegraphics[scale=0.47]{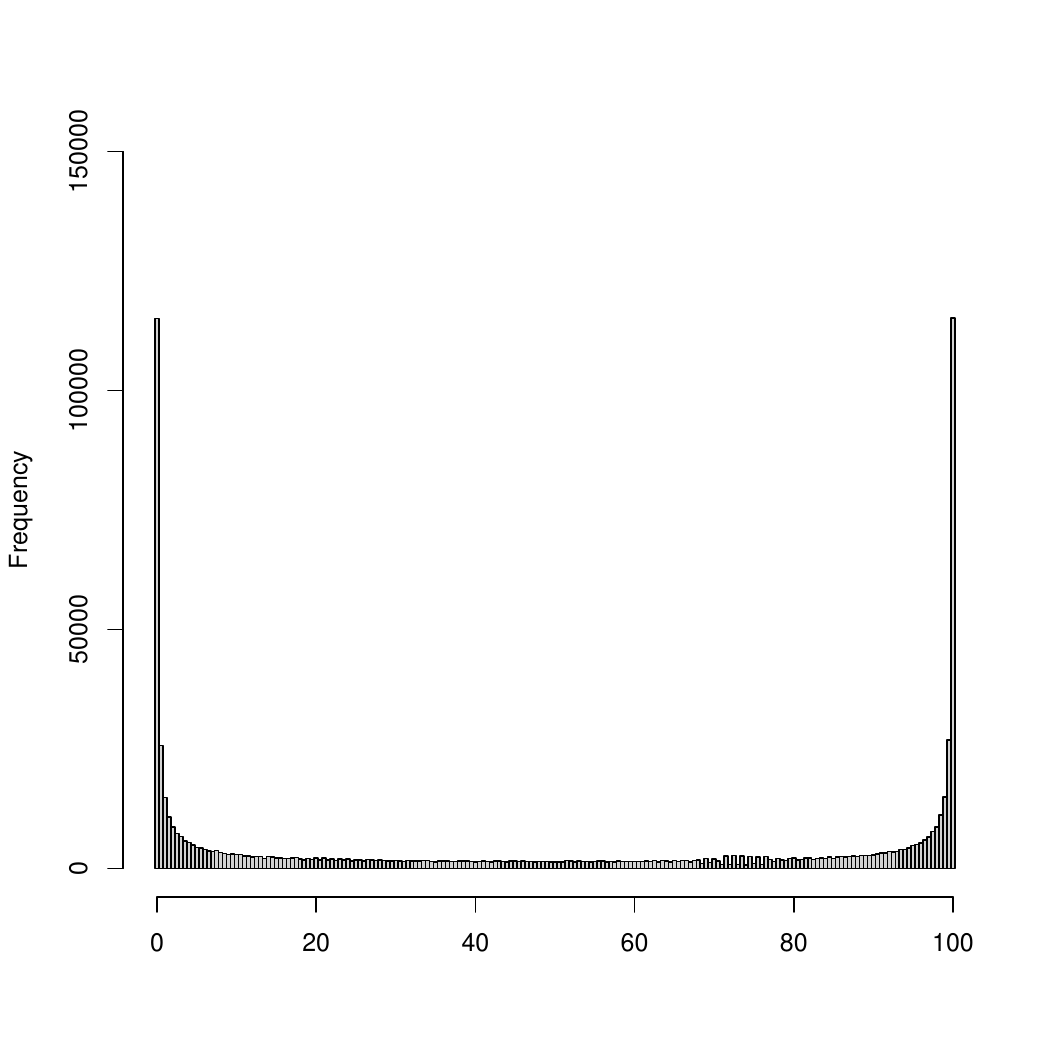}}\;
\subfloat[][\label{fig:histispd_fcm}]{\includegraphics[scale=0.47]{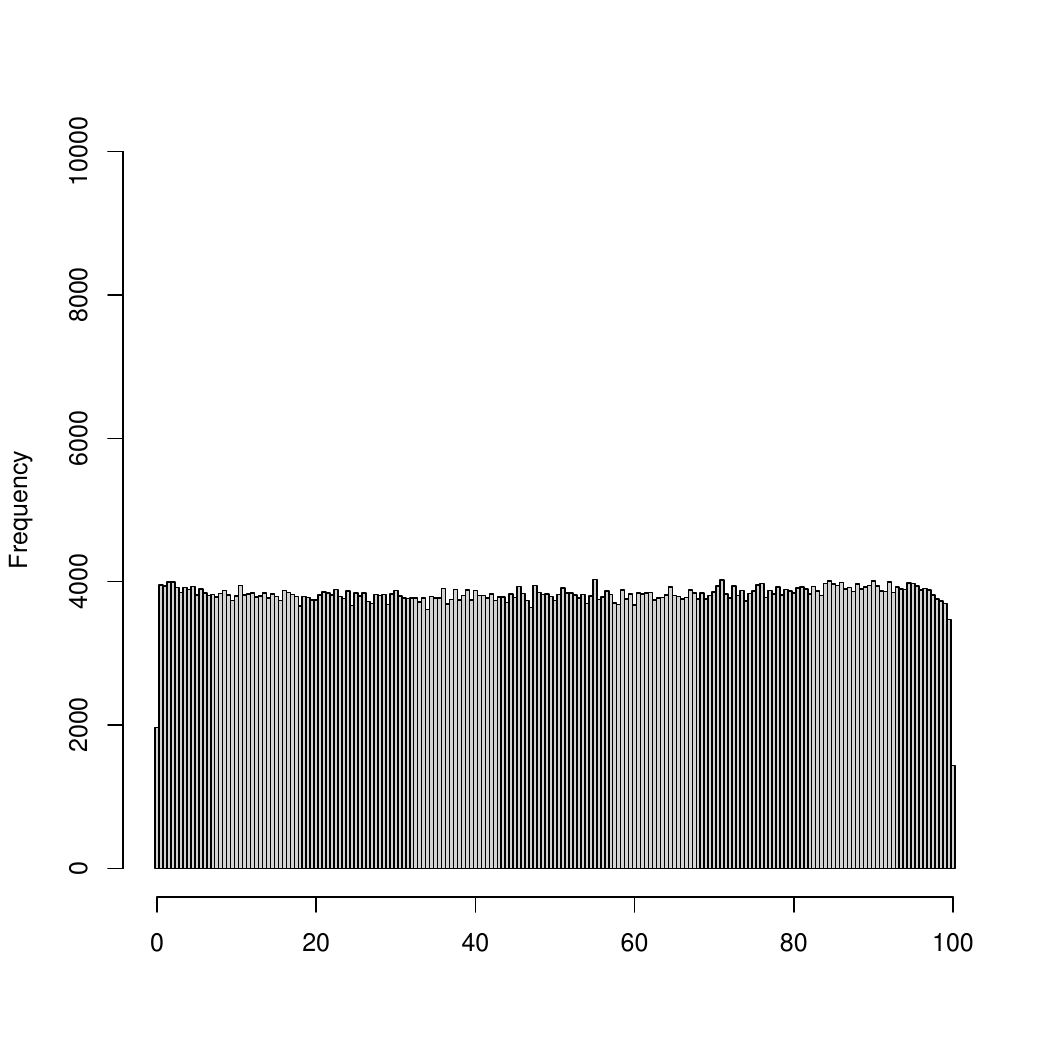}}
}
\caption{Empirical histograms of (a) $\text{ISPD}$ and (b) $\text{ISPD}^{\textsc{fcm}}$ for the medium perturbation scenario (all departments and datasets combined).}
\label{fig:histispd}
\end{figure}


\begin{figure}[tb]
\centering{
\subfloat[][\label{fig:histscaledstdmin}]{\includegraphics[scale=0.47]{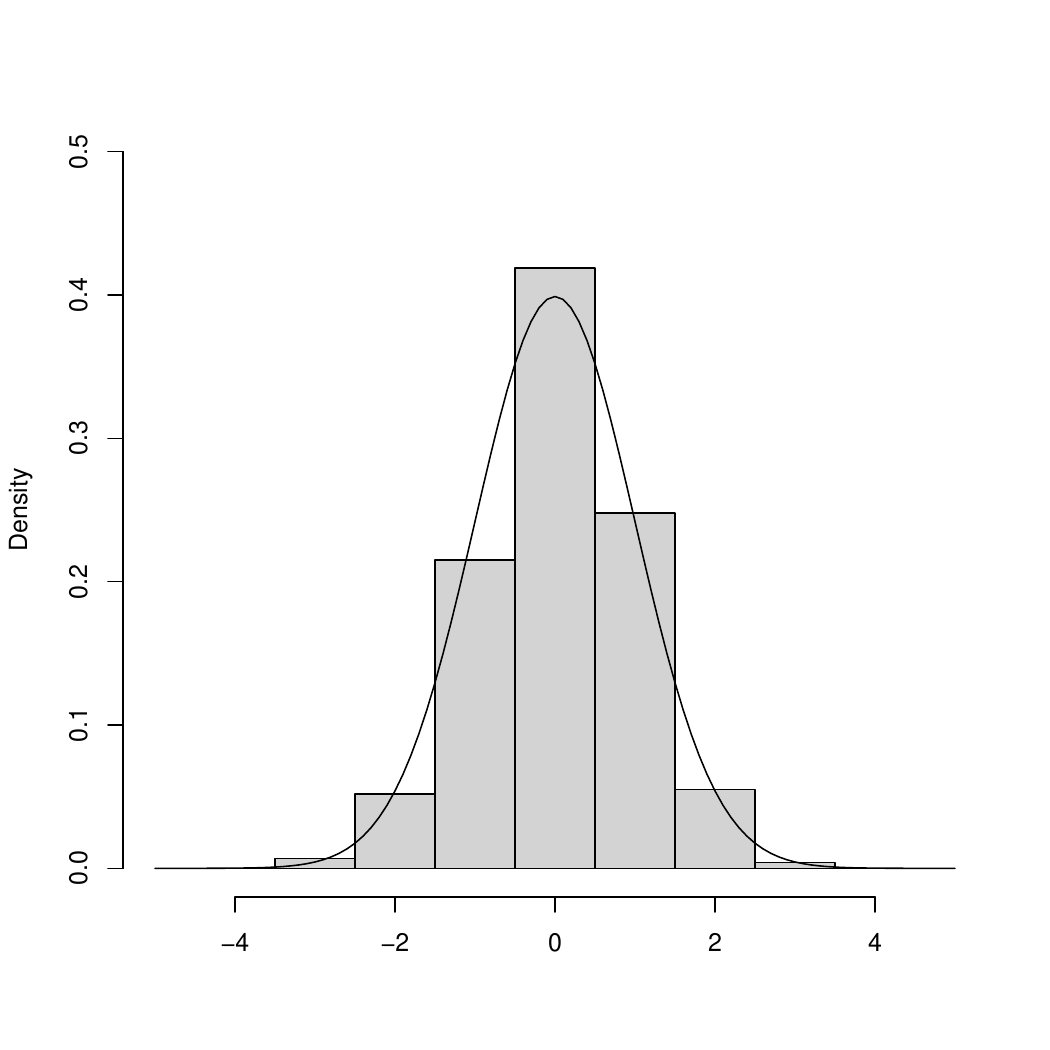}}\;
\subfloat[][\label{fig:histscaledstdmax}]{\includegraphics[scale=0.47]{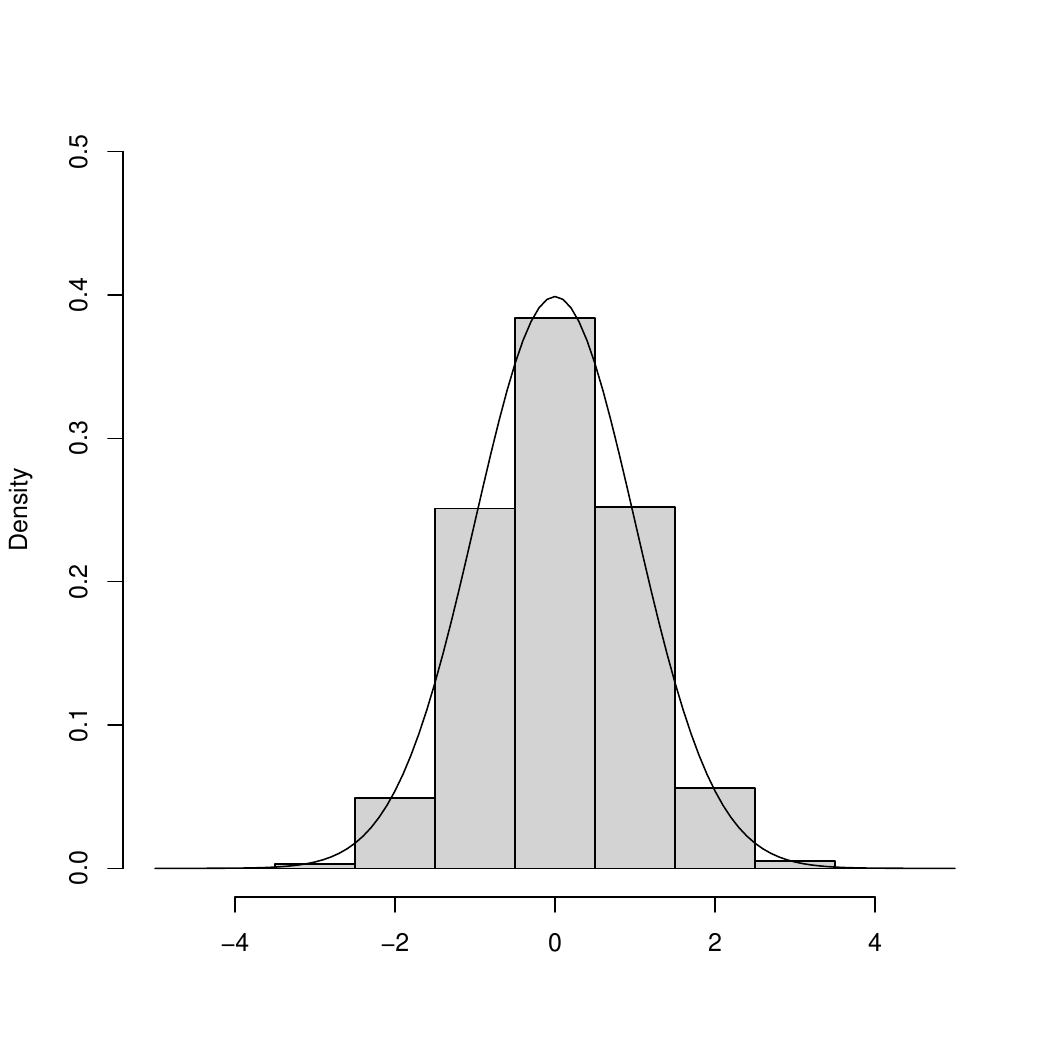}}
}
\caption{Empirical histograms of scaled averages (divided by the corresponding theoretical standard deviation) for the minimum (a) and maximum (b) size departments.}
\label{fig:histscaledstd}
\end{figure}


Table~\ref{tab:sims} reports summary statistics of the empirical MAD and PDC distributions for the four competing indices under the four perturbation levels. With respect to MAD, the original $\text{ISPD}$ performs poorly in all scenarios, with values ranging between 12.75 and 14.63. This reflects its well-known tendency to polarization, whereby several departments receive identical scores despite underlying performance differences. A similar pattern, albeit with smaller magnitudes (8.73–10.09), emerges for $\text{ISPD}^{\textsc{np}}$: the instability of the non-parametric estimator $\hat{\rho}_d^{\textsc{np}}$, based on second-order statistics, translates into substantial variability and ranking distortions. In contrast, $\text{ISPD}^{\textsc{fcm}}$ and $\text{ISPD}^{\textsc{rim}}$ exhibit dramatically lower MAD values (0.02–2.27), with the former consistently achieving the best performance. As expected, their MAD increases with the perturbation level, although the deterioration remains limited (in this respect, no systematic trend is observed for $\text{ISPD}$ or $\text{ISPD}^{\textsc{np}}$). The PDC results mirror these findings: $\text{ISPD}^{\textsc{fcm}}$ and $\text{ISPD}^{\textsc{rim}}$ substantially outperform the original $\text{ISPD}$ and ISPD$^{\textsc{np}}$, with the former slightly dominating the latter. Notably, $\text{ISPD}^{\textsc{np}}$ emerges as the worst-performing index according to PDC. Again, this reflects the instability of $\hat{\rho}_d^{\textsc{np}}$, which more frequently induces pairwise rank reversals.

Finally, to provide some more details, for the medium perturbation scenario Figure~\ref{fig:histispd} shows the Montecarlo distributions of ISPD in~\eqref{GEM1} and $\text{ISPD}^{\textsc{fcm}}$, i.e., for the worst and the best index according to the MAD statistics. In particular, the distribution of $\text{ISPD}^{\textsc{fcm}}$ is quite uniform, as expected, thus assuring a better and fairer ranking of Departments than the one provided by ISPD in~\eqref{GEM1}. Very similar behaviors of the above Montecarlo distributions are observed for the other scenarios. At last, for the department with the minimum (24) and the maximum (464) size Figure~\ref{fig:histscaledstd} depicts the empirical histogram (across the 1000 replications) of scaled averages (divided by the theoretical standard deviation). Even for the smallest department, deviations from Normality are not relevant, thereby confirming that the modified version of CLT discussed in Section~\ref{sec:estpubl} applies.


\section{Conclusions}\label{sec:concl}

In this paper, we examine the methodological features of performance-based funding systems for academic departments. In particular, we focus on the Italian case and on the Standardized Department Performance Index (in Italian, ISPD), which underlies the construction of department rankings and the subsequent allocation of funds. The ISPD is computed as a function of the scaled average of standardized scores assigned to a set of Scientific Products (SPs) submitted by each department, where standardization is performed at the level of Scientific Disciplinary Sectors (SDSs). While this procedure aims to ensure comparability across heterogeneous fields, it implicitly assumes that standardized scores are independent within departments. 

We claim that the above assumption is statistically inappropriate, since SPs within the same department exhibit intra-departmental correlation due to shared research environments, institutional strategies, recruitment patterns, and disciplinary specialization. Ignoring this within-department homogeneity leads to an underestimation of the variances of the benchmark distributions for scaled averages and, consequently, to an artificial compression and polarization of final ISPD scores. This is confirmed by empirical evidence from the 2017 and 2022 rankings, showing that a large number of departments receive either the maximum or minimum score. As a consequence, unrealistic ties at the tails of the distribution are generated, which undermine the discriminatory power of the index.

To address the above issue, we propose a model where the intra-departmental correlation is linked to department size. Such a model requires a limited data burden, since it relies on department-level rather than SP-level data, that is, on macro- rather than micro-data. Model parameters can be estimated via standard Maximum Likelihood (ML) techniques, thereby allowing for the correct recovery of the benchmark variances and, in turn, the proper adjustment of the ISPD. The whole ML estimation framework has also been adapted to achieve estimation of the correlation model parameters from publicly available ISPD data, which are affected by coarsening due to rounding and/or truncation. To this end, a novel probability distribution, termed \emph{Betoidal}, has been introduced.

Estimation results suggest the presence of varying intra-departmental correlation levels for both the 2017 and 2022 ISPD data. In detail, within-department correlation tends to decrease with size, in line with subject-matter expectations. These findings confirm the need for substantial adjustments to the ISPD. Evidence from the 2017 data also forms the basis for the design of a simulation study, where standardized scores are generated according to the postulated intra-departmental correlation model as well as to three perturbed versions of it. The study highlights that the model-adjusted ISPD is always closer to the ground truth, not only with respect to the original index, but also to other competing proposals relying on SP-level data.

Beyond the technical corrections, our findings raise broader issues of accountability and transparency in research evaluation systems. When rankings are directly linked to substantial funding allocations, the statistical properties of the underlying indicators become matters of public accountability. An index that systematically inflates ties or compresses variance is likely to distort funding decisions, weaken the legitimacy of evaluation outcomes, and reduce trust among evaluated institutions. Moreover, transparency requires that evaluation agencies explicitly state the statistical assumptions embedded in the adopted performance indicators, including assumptions about the underlying correlation structure. When intra-department correlation is structurally present, failing to properly model it may generate systematic bias in favor of specific organizational configurations (e.g., departments with particular size or internal cohesion patterns). In this sense, our analysis shows that methodological choices in index construction are not neutral technical details, but elements with concrete distributive consequences.

Several lines of future research emerge from this work. First, further methodological developments might be needed to keep the correlation structure postulated by the model tailored to the one actually implied by the real-world dynamic, also in light of the quite frequent changes to the evaluation protocol. Furthermore, related approaches could point to data from other national evaluation systems, assessing whether similar issues arise in different institutional contexts. Finally, a broader perspective should examine how choices in the statistical design of evaluation metrics affect long-term dynamics in academic systems (including the correlation structure), particularly under performance-based funding regimes. In conclusion, this paper shows that properly accounting for intra-departmental correlation is not only a statistical refinement but a necessary condition for ensuring fairness, transparency, and accountability in research evaluation systems. When rankings are tied to substantial financial resources, methodological rigor in the construction of the adopted indicators becomes a central component of sound research governance.


\bibliographystyle{abbrvnat}      
\bibliography{Biblio_Betoidal}

\appendix


\section{Central Limit Theorem for correlated observations}\label{app:clt}

Consider a Virtual Department with $N$ scientific products. Let $Z_i$, $i\in \mathcal{N}=\{1,\dots,N\}$, denote the random standardized score associated with product $i$, with $E(Z_i)=0$ and $V(Z_i)=1$. The variance of the scaled average $\tilde{Z}=\sum_{i\in \mathcal{N}} Z_i  /\sqrt{N}$
is given by
\[
V(\tilde{Z})
= \frac{1}{N}\sum_{i\in\mathcal{N}}V(Z_i)
  + \frac{1}{N}\sum_{i\in \mathcal{N}}\sum_{i'\in \mathcal{N}\setminus i}\text{Cov}(Z_i,Z_{i'})
= \frac{1}{N}\left(N+\sum_{i\in \mathcal{N}}\sum_{i'\in \mathcal{N}\setminus i}\rho_{ii'}\right),
\]
where $\rho_{ii'}=\text{Corr}(Z_i,Z_{i'})$. Letting 
\[
\rho=\frac{1}{N(N-1)}\sum_{i\in \mathcal{N}}\sum_{i'\in \mathcal{N}\setminus i}\rho_{ii'}
\]
be the average pairwise intra-departmental correlation, we have $V(\tilde{Z})= 1 + (N-1)\rho$.
Hence, unless $\rho$ decreases sufficiently fast as $N$ increases, the variance of the scaled average does not stabilize. In particular, if $\rho$ is constant and positive, ${\rm V}(\tilde{Z})$ diverges linearly in $N$, reflecting the well-known effect of dependence on aggregation.

A Central Limit Theorem (CLT) for $\tilde{Z}$ continues to hold under suitable weak-dependence conditions. In detail, a sufficient requirement is that $N\rho \longrightarrow \gamma \quad \text{as } N \to \infty$, for some finite constant $\gamma \ge 0$. 
Equivalently, $\rho = O(1/N)$. Under this condition,
\[
\frac{\tilde{Z}}{\sqrt{1+\gamma}}
\;\xrightarrow{d}\;
\mathcal{N}(0,1).
\]
This result follows from standard central limit theorems for arrays of weakly dependent random variables (see, e.g.,~\citealt{Chudiketal2011} as well as classical treatments such as~\citealt{DavydovIbragimov1971} and~\citealt{HallHeyde1980}). 

In the context of this paper, the correlation function defined in~\eqref{eq:rhotheta} implies, when $\beta<0$, that pairwise correlations decay sufficiently fast with departmental size, thereby satisfying the weak-dependence requirement and ensuring asymptotic normality of the standardized scaled average.


\section{Details on the Betoidal distribution}\label{app:moments}

Since both the error function $\mathrm{erf}(\cdot)$ and its inverse pass through the origin, it follows directly from~\eqref{eq:qfunctbet} that $F_X^{-1}(0.5;\sigma)=1/2$ for every $\sigma>0$, hence the median of $X\sim \text{Betoidal}(\sigma)$ is always equal to $0.5$.

To show that $E(X)=0.5$, it is sufficient to prove that the density in~\eqref{eq:pdf} is symmetric around $0.5$, with symmetry requiring
\[
f_X(0.5+x;\sigma)=f_X(0.5-x;\sigma).
\]
Using the explicit form of the PDF, this condition reduces to
\[
\phi\bigl(\sqrt{2}\,\mathrm{erf}^{-1}(x)\bigr)
\exp\{[\mathrm{erf}^{-1}(x)]^2\}
=
\phi\bigl(\sqrt{2}\,\mathrm{erf}^{-1}(-x)\bigr)
\exp\{[\mathrm{erf}^{-1}(-x)]^2\},
\]
which holds because $\mathrm{erf}^{-1}(-x)=-\mathrm{erf}^{-1}(x)$ and $\phi(x)=\phi(-x)$. 
Therefore, the distribution is symmetric around $0.5$, implying $E(X)=0.5$. An equivalent derivation exploits the representation
\[
X=\Phi(\tilde{Z}), 
\qquad \tilde{Z}\sim N(0,\sigma^2).
\]
Then
\[
E(X)
=
\int_{-\infty}^{+\infty}\Phi(z)\frac{1}{\sigma}
\phi\!\left(\frac{z}{\sigma}\right)\!dz.
\]
Splitting the integral at zero and using classical results on Gaussian integrals~\citep{Owen1980} yields
$E(X)=1/2$.

To derive the variance, it is useful to rely on the transformation $k=z/\sigma$ to obtain
\[
\begin{split}
E(X^2)
&=
\int_{-\infty}^{+\infty}
\{\Phi(z)\}^2
\frac{1}{\sigma}
\phi\!\left(\frac{z}{\sigma}\right)\!dz \\
&=
\int_{-\infty}^{+\infty}
\{\Phi(k\sigma)\}^2
\phi(k)\,dk
=
\frac{1}{\pi}
\atan\!\bigl(\sqrt{1+2\sigma^2}\bigr),
\end{split}
\]
where the last equality follows from identities reported in \cite{Owen1980}. Equivalently,
\[
E(X^2)
=
\frac12 - 2T\!\left(0,\frac{1}{\sqrt{1+2\sigma^2}}\right),
\]
where $T(\cdot,\cdot)$ denotes Owen's $T$ function. The variance in~\eqref{eq:varbetoidal} follows immediately from $V(X)=E(X^2)-1/4$.

For ML estimation, let $\bm{x}=(x_1,\dots,x_n)^\top$ be i.i.d. draws from $\text{Betoidal}(\sigma)$. 
Using the change-of-variable representation $x_i=\Phi(\tilde{z}_i)$ with $\tilde{z}_i\sim N(0,\sigma^2)$, the log-likelihood can be written as
\[
\ell_X(\sigma;\bm{x})
=
n\log\frac{\sqrt{2\pi}}{\sigma}
+
\sum_{i=1}^n
\log\phi\!\left(\frac{\Phi^{-1}(x_i)}{\sigma}\right)
+
\sum_{i=1}^n
[\mathrm{erf}^{-1}(2x_i-1)]^2.
\]
The first and second derivatives with respect to $\sigma$ are
\begin{equation}\label{eq:deriv}
\frac{\partial \ell_X}{\partial\sigma}
=
-\frac{n}{\sigma}
+
\frac{\sum_{i=1}^n[\Phi^{-1}(x_i)]^2}{\sigma^3},
\qquad
\frac{\partial^2 \ell_X}{\partial\sigma^2}
=
\frac{n}{\sigma^2}
-
\frac{3\sum_{i=1}^n[\Phi^{-1}(x_i)]^2}{\sigma^4}.
\end{equation}
Setting the score equal to zero gives
\[
s
=
\sqrt{\frac{1}{n}
\sum_{i=1}^n
[\Phi^{-1}(x_i)]^2},
\]
which is the unique solution of 
$\partial \ell_X/\partial\sigma=0$. 
Evaluating the second derivative at $\sigma=s$ yields
\[
\left.
\frac{\partial^2 \ell_X}{\partial\sigma^2}
\right|_{\sigma=s}
<0,
\]
thus confirming that $s$ is the maximum likelihood estimate.

The expected Fisher information is
\[
\mathcal{I}_X(\sigma)
=
-nE\!\left(
\frac{\partial^2}{\partial\sigma^2}
\log f_X(X;\sigma)
\right)
=
\frac{2n}{\sigma^2}.
\]
Denoting by $S$ the ML estimator, we then have
\[
V(S)=\mathcal{I}_X(\sigma)^{-1}=\frac{\sigma^2}{2n}.
\]
Since the Betoidal distribution is obtained via the monotone transformation
\[
X=\Phi(\tilde{Z}),
\qquad \tilde{Z}\sim N(0,\sigma^2),
\]
ML estimation of $\sigma$ is equivalent to estimating the scale parameter of a zero-mean Normal sample $\{\tilde{Z}_i\}$, where $\tilde{Z}_i=\Phi^{-1}(X_i)$. Hence,
\[
S=
\sqrt{\frac{1}{n}\sum_{i=1}^n \tilde{Z}_i^2}
\]
coincides with the well-known ML estimator of the standard deviation in the Normal model. The asymptotic variance $\sigma^2/(2n)$ is therefore consistent with classical Normal theory.


\section{First and second derivatives}\label{app:nr}

The score and Hessian contributions in Section~\ref{subsec:coarse} are
\[
s_{\iota,dj}^{(\alpha)}(\bm{\theta})
= \frac{\eta_{dj}(\bm{\theta})}{\pi_{dj}(\bm{\theta})}
\qquad\text{and}\qquad
h_{\iota,dj}^{(\alpha)}(\bm{\theta})
= \frac{\tau_{dj}(\bm{\theta})\pi_{dj}(\bm{\theta})
       -[\eta_{dj}(\bm{\theta})]^2}
       {[\pi_{dj}(\bm{\theta})]^2},
\]
where $\pi_{dj}(\bm{\theta})$ is defined in~\eqref{eq:pd} and
\[
\begin{split}
\eta_{dj}(\bm{\theta}) 
&= \frac{(N_d-1)\delta_d^{(\alpha)}}{2\sqrt{\pi}\sigma_d^3}
\biggl\{
\exp\{-[e(k_{j-})/\sigma_d]^2\}e(k_{j-})
-
\exp\{-[e(k_{j+})/\sigma_d]^2\}e(k_{j+})
\biggr\} \\
\tau_{dj}(\bm{\theta}) 
&=
\frac{\tilde{N}-\exp\bigl\{\alpha+\beta(N_d-1)\bigr\}}
     {\tilde{N}+\exp\bigl\{\alpha+\beta(N_d-1)\bigr\}}
\,\eta_{dj}(\bm{\theta}) 
+\frac{(N_d-1)\delta_d^{(\alpha)}}{2\sigma_d^2}
\biggl\{
q_{dj}(\bm{\theta}) - 3\eta_{dj}(\bm{\theta})
\biggr\},
\end{split}
\]
with
\[
q_{dj}(\bm{\theta})
=
\frac{(N_d-1)\delta_d^{(\alpha)}}{\sqrt{\pi}\sigma_d^5}
\biggl\{
\exp\{-[e(k_{j-})/\sigma_d]^2\}\bigl[e(k_{j-})\bigr]^3
-
\exp\{-[e(k_{j+})/\sigma_d]^2\}\bigl[e(k_{j+})\bigr]^3
\biggr\}.
\]
For the truncated case (Section~\ref{subsec:coarsetrunc}), the derivatives become
\[
s_{\iota^{\star},dj}^{(\alpha)}(\bm{\theta},\tilde{s}_{J^{\star}-})
=
\frac{\eta^{\star}_{dj}(\bm{\theta},\tilde{s}_{J^{\star}-})}
     {\pi^{\star}_{dj}(\bm{\theta},\tilde{s}_{J^{\star}-})}
\qquad\text{and}\qquad
h_{\iota^{\star},dj}^{(\alpha)}(\bm{\theta},\tilde{s}_{J^{\star}-})
=
\frac{\tau^{\star}_{dj}(\bm{\theta},\tilde{s}_{J^{\star}-})
      \pi^{\star}_{dj}(\bm{\theta},\tilde{s}_{J^{\star}-})
      -[\eta^{\star}_{dj}(\bm{\theta},\tilde{s}_{J^{\star}-})]^2}
     {[\pi^{\star}_{dj}(\bm{\theta},\tilde{s}_{J^{\star}-})]^2},
\]
where $\pi^{\star}_{dj}(\bm{\theta},\tilde{s}_{J^{\star}-})$ is given in~\eqref{eq:pstard} and
\[
\begin{split}
\eta^{\star}_{dj}(\bm{\theta},\tilde{s}_{J^{\star}-})
&=
\frac{\eta_{dj}(\bm{\theta})
-
\pi^{\star}_{dj}(\bm{\theta},\tilde{s}_{J^{\star}-})
\,\omega^{\star}_d(\bm{\theta},\tilde{s}_{J^{\star}-})}
     {1-F_X(\tilde{s}_{J^{\star}-};\sigma_d)} \\
\tau^{\star}_{dj}(\bm{\theta},\tilde{s}_{J^{\star}-})
&=
\frac{
\tau_{dj}(\bm{\theta})
-2\eta^{\star}_{dj}(\bm{\theta},\tilde{s}_{J^{\star}-})
 \omega^{\star}_d(\bm{\theta},\tilde{s}_{J^{\star}-})
-\pi^{\star}_{dj}(\bm{\theta},\tilde{s}_{J^{\star}-})
 \xi^{\star}_d(\bm{\theta},\tilde{s}_{J^{\star}-})
}
{1-F_X(\tilde{s}_{J^{\star}-};\sigma_d)},
\end{split}
\]
with
\[
\begin{split}
\omega^{\star}_d(\bm{\theta},\tilde{s}_{J^{\star}-})
&=
\frac{\partial}{\partial\alpha}
\{1-F_X(\tilde{s}_{J^{\star}-};\sigma_d)\} \\
&=
\frac{(N_d-1)\delta^{(\alpha)}_d}
     {2\sqrt{\pi}\sigma^3_d}
\exp\{-[e(\tilde{s}_{J^{\star}-})/\sigma_d]^2\}
\,e(\tilde{s}_{J^{\star}-}) \\
\xi^{\star}_d(\bm{\theta},\tilde{s}_{J^{\star}-})
&=
\frac{\partial}{\partial\alpha}
\omega^{\star}_d(\bm{\theta},\tilde{s}_{J^{\star}-}) \\
&=
\omega^{\star}_d(\bm{\theta},\tilde{s}_{J^{\star}-})
\Biggl\{
\frac{\tilde{N}-\exp\bigl\{\alpha+\beta(N_d-1)\bigr\}}
     {\tilde{N}+\exp\bigl\{\alpha+\beta(N_d-1)\bigr\}}
+
\frac{(N_d-1)\delta^{(\alpha)}_d}
     {2\sigma^4_d}
\bigl[
2\{e(\tilde{s}_{J^{\star}-})\}^2
-3\sigma^2_d
\bigr]
\Biggr\}.
\end{split}
\]


\section{Generation of correlated standardized scores}\label{app:simsdgp}

The following scheme is adopted to generate $N_d$ standardized scores for department $d$, with an intra-departmental correlation level close to a target value $\rho_d$. Starting from the standardized distribution introduced in Section~\ref{sec:simulazio}, we proceed as follows. First, $M_d$ independent standardized scores are drawn and each of them is replicated $k_d$ times, so that $M_d k_d \leq N_d$. Then, one additional independent score is drawn and replicated $\check{k}_d < k_d$ times, with $0 \leq \check{k}_d \leq N_d - M_d k_d$.
Finally, $N_d - M_d k_d - \check{k}_d$ further independent draws are added in order to complete the set of $N_d$ standardized scores. The total number of independent draws is therefore $N_d^{\star} = N_d - M_d (k_d - 1) - (\check{k}_d - 1)$, and these draws can be collected in the vector $\bm{z}_d^* = (z_{d,1}^*, \dots, z_{d,N_d^{\star}}^*)^\top$. In this setting, the scaled average $\tilde{z}_d$ can be written as
\[
\tilde{z}_d = \frac{1}{\sqrt{N_d}} 
\left\{
\sum_{i=1}^{M_d} k_d z_{d,i}^*
+ \check{k}_d z_{d,M_d+1}^*
+ \sum_{i=M_d+2}^{N_d^{\star}} z_{d,i}^*
\right\}.
\]
Let $\tilde{Z}_d$ denote the corresponding random variable. Since the components of $\bm{z}_d^*$ are independent with zero mean and unit variance, it follows that $\tilde{Z}_d$ has zero mean and variance
\[
\sigma_d^2
= 1 + \frac{M_d k_d (k_d - 1) + \check{k}_d (\check{k}_d - 1)}{N_d}.
\]
Moreover, as $N_d^{\star} \to \infty$, the distribution of $\tilde{Z}_d$ is asymptotically Normal by the Lindeberg–Feller CLT, since $\tilde{z}_d$ is a normalized sum of independent (though not identically weighted) random variables satisfying the Lindeberg condition.

In light of~\eqref{eq:vrho}, the implied intra-departmental correlation can be written as
\begin{equation}
\label{eq:rhomk}
\rho_d
= \frac{M_d k_d (k_d - 1) + \check{k}_d (\check{k}_d - 1)}
{N_d (N_d - 1)}.
\end{equation}
Equation~\eqref{eq:rhomk} shows that, in principle, any target correlation level $\rho_d$ can be obtained through a suitable choice of $M_d$, $k_d$ and $\check{k}_d$. To select a triplet $(M_d, k_d, \check{k}_d)$ approximating the desired $\rho_d$ in~\eqref{eq:rhotheta}, we adopt the following heuristic algorithm. As a starting point, set $\check{k}_d = 0$. From~\eqref{eq:rhomk} we obtain
\[
M_d = \frac{\rho_d N_d (N_d - 1)}{k_d (k_d - 1)}.
\]
The feasibility condition $M_d k_d \leq N_d$ implies $\rho_d \leq (k_d - 1)/(N_d - 1)$, that is, $k_d \geq 1 + \rho_d (N_d - 1)$. Since both $k_d$ and $M_d$ must be integers, a rounding scheme is required. Specifically:
\begin{itemize}
\item $k_d$ is set equal to the smallest integer $k_d^{(r)}$ such that $k_d^{(r)} \geq 1 + \rho_d (N_d - 1)$;

\item given $k_d^{(r)}$, $M_d$ is set equal to the largest integer $M_d^{(r)}$ such that
\[
M_d^{(r)} \leq 
\frac{\rho_d N_d (N_d - 1)}
{k_d^{(r)} (k_d^{(r)} - 1)}.
\]
\end{itemize}
This choice maximizes $M_d$ under the integer constraint.

If $M_d^{(r)} k_d^{(r)} < N_d$, the rounding implies an actual intra-departmental correlation $\rho_d^{(r)} \leq \rho_d$. To compensate for this downward bias, an additional cluster of size $\check{k}_d < k_d$ is introduced.
The required value of $\check{k}_d$ is obtained as the positive solution $\check{k}_d^s$ of the quadratic equation
\[
\check{k}_d^2 - \check{k}_d
- \Bigl(\rho_d - \rho_d^{(r)}\Bigr) N_d (N_d - 1) = 0,
\]
namely
\[
\check{k}_d^s
= \frac{1}{2}
+ \sqrt{
\frac{1}{4}
+ \Bigl(\rho_d - \rho_d^{(r)}\Bigr) N_d (N_d - 1)
}.
\]
At last, $\check{k}_d$ is set equal to the smallest integer between $N_d - M_d^{(r)} k_d^{(r)}$ and the integer closest to $\check{k}_d^s$, so as to avoid generating more than $N_d$ observations. With this algorithm, the achieved intra-departmental correlations exhibit a small difference from the target ones (typically less than 5\% in relative terms).

\end{document}